%% file: main.tex
\documentclass{article}
\usepackage{arxiv}
\usepackage[utf8]{inputenc}
\input{packages_commands}
\usepackage{esint}
\hypersetup{
     colorlinks   = true,
     linkcolor    = blue,
     citecolor    = blue,
     urlcolor     = blue
}
\title{Topology optimization of permanent magnets for stellarators}
\author[ ]{Caoxiang Zhu$^*$, Kenneth Hammond, Thomas Brown, David Gates, Michael Zarnstorff, Keith Corrigan, Marc Sibilia, Eliot Feibush}
\affil[ ]{Princeton Plasma Physics Laboratory, Princeton University, P.O. Box 451, New Jersey 08543, USA}
\affil[*]{\textit{Email: czhu@pppl.gov}}

\begin{document}
\maketitle
%%%%%%%%%%%%%%%%%%%%%%%%%%%%%%%%%%%%%%%%%%%%%%%%%%%%%%%%%%%%%%%%%%%%%%%%%%%%%%%%
% \documentclass{iopart}
% \input{packages_commands}
% \usepackage{esint}
% %\usepackage{hyperref}
% %\usepackage{natbib}
% %\usepackage{doi}
% \hypersetup{
%      colorlinks   = true,
%      linkcolor    = blue,
%      citecolor    = blue,
%      urlcolor     = blue
% }
% \hyphenation{FAMUS}

% \begin{document}
% \title{Topology optimization of permanent magnets for stellarators}
% \author{Caoxiang Zhu, Kenneth Hammond, Thomas Brown, David Gates, Michael Zarnstorff, Keith Corrigan, Marc Sibilia and Eliot Feibush}
% \address{Princeton Plasma Physics Laboratory, Princeton University, P.O. Box 451, New Jersey 08543, USA}
% \ead{\mailto{czhu@pppl.gov}}
% %\keyword{stellarator, permanent magnet, optimization, topology optimization, simple coil, design}

%\ioptwocol
\begin{abstract}
We introduce a topology optimization method to design permanent magnets for advanced stellarators. 
Recent researches show that permanent magnets have great potentials to simplify stellarator coils.
We adopt state-of-the-art numerical techniques to determine the presence of magnets in the entire designing space.
The FAMUS code is developed and it can design engineering-feasible permanent magnets for general stellarators satisfying the constraints of the maximum material magnetization and explicit forbidden regions.
FAMUS has been successfully verified against the previously proposed linear method. 
Three different permanent magnet designs together with planar TF coils for a half-Tesla NCSX configuration have been obtained for demonstrations.
The designs have good accuracy in generating the desired equilibrium and offer considerably large plasma access on the outboard side.
The results show that FAMUS is a flexible, advanced numerical tool for future permanent magnet stellarator designs.
\end{abstract}

\section{Introduction}
Stellarators have attractive properties, like low recirculating power, steady-state operation and free of disruptions, but they also suffer the engineering complexities, especially complicated non-planar coils.
Recently, the concept of using permanent magnets to simplify stellarator coils was introduced.
Helander \etal \cite{HelanderPM} proposed a mathematical solution of a curl-free, ``one-sided" magnetization distribution such that the effective surface current can be determined by the NESCOIL \cite{NESCOIL} method.
Landreman \cite{REGCOIL_PM} developed a least-squares minimization method to optimize the magnetization in space together with a Picard iteration method to constrain the maximum magnetization.
In a previous paper \cite{ZhuPM01}, we introduced a linear method to design perpendicular magnets which employs the idea that current potential is the magnetic dipole moment per unit area.
An iterative scheme was also developed to incrementally stack multi-layer magnetizations using the linear method.
%However, both methods can not embrace engineering constraints explicitly.
However, to obtain engineering feasible designs, we need more sophisticated methods.

Stellarator optimization is normally divided into two steps: 1) configuration optimization using fixed-boundary MHD equilibrium codes to obtain desirable physics properties \cite{STELLOPT, ROSE}; 2) coil optimization as a ``reverse-engineering'' step to design coils for supporting the target equilibrium \cite{REGCOIL, FOCUS00}.
The two steps usually will have to be iterated several times until a good balanced solution is obtained.
Afterwards, a detailed engineering design will be carried out.
The design of permanent magnets is an alternative approach to the coil optimization step.
Therefore, in this problem, we are assuming there exists a target configuration and only some engineering constraints will be considered.
After an applicable design was obtained, more precise engineering analyses would be performed.

We identify the problem of designing permanent magnets as a topology optimization problem.
Topology optimization \cite{topology_optimization} is defined as a mathematical method to determine material layout within a given design space for a maximum performance subjected to some constraints.
It was first utilized in structural mechanics \cite{Rozvany2001} and later was commonly used in electromagnetics \cite{Campelo2010}.
To efficiently determine the material layout, there are several numerical techniques developed, including the homogenization method \cite{Bendsoe1988}, the density method \cite{Dyck1996}, the ON/OFF method \cite{Anagnostou1992}, the level-set method \cite{Zhou2010} and others.
Each technique has its own advantages and drawbacks.
For the problem of designing permanent magnets for stellarators, we are going to employ the density method because of simplicity, which shall be briefly introduced later in this paper.
While interesting, other techniques are outside of the scope of this work, and the reader is referred to the bibliography for further details.

The forward problem for the optimization is to calculate the magnetic field produced by a given magnetization.
In textbooks \cite{JacksonCED}, there are three ways, magnetic charges, effective currents and discretized magnetic dipoles.
To numerically compute the magnetic field, various methods \cite{Abert2013} have been developed using the finite-element method, the finite-difference method and others.
The problem that we are going to discuss in this paper is distinct from other existing applications.
First, {the computation domain in this problem} is much larger.
A stellarator usually has a major radius bigger than 1 meter (some even larger than 10 meters).
This indicates that a large amount of magnetic material is required.
For example, a half-Tesla three-period quasi-axisymmetric stellarator might need 3 tons of magnets \cite{ZhuPM01}.
Second, the geometry is more complicated.
For modern stellarators, the configurations are optimized to have desirable properties, like good particle confinement and MHD stability.
Therefore, the plasma shape is generally complicated and so is the vessel.
The complicated geometries create difficulty when generating meshes in the design space.
Third, the distance between the plasma and permanent magnets is normally sizable.
Permanent magnets are sensitive to temperature and might be demagnetized under neutron fluxes \cite{Alderman2002}.
Thus, magnets need to be placed outside the vessel (and outside the blanket for fusion reactors).
%Problems of designing permanent magnets for stellarators are different from the others.
%The magnets should be placed outside the vessel, which means the distance between the magnets and plasma is non-trivial. 
%In addition, we need a large amount of magnets, while 
Because of all the reasons mentioned above, we use the dipole representation to calculate the magnetic field from magnetizations in this paper.
Elements that are sufficiently small will be approximated by one magnetic dipole.
The detailed distribution inside each element will not be important during the stage of optimization.
In addition, the present magnets are usually homogeneously magnetized which could be well approximated by a dipole when evaluating the far field.
The dipole representation is concise and easy, while the accuracy increases with distance.
According to Ref. \cite{Petruska2013}, the relative error of the dipole approximation for a cubical homogeneous permanent magnet is about 0.01\% at the distance of 7 times of the magnet's dimension.

This paper is organized as follows.
The methodology of designing permanent magnets using the topology optimization method is described in section \ref{method}.
Section \ref{benchmark} shows a benchmark with the previous linear method for validations.
In section \ref{application}, three designs of permanent magnets for a half-Tesla three-period quasi-axisymmetric stellarator are demonstrated.
It includes a perpendicular only solution, an orientation-optimized solution and a solution combined with simplified geometry.
We then summarize in section \ref{summary}.

\section{Methodology} \label{method}
\subsection{Magnetic field calculation}

The magnetic field generated by a magnetic dipole is calculated by the Biot-Savart law.
%\begin{equation} \label{eq:dipole}
%    \vect{B} = \frac{\mu_0}{4\pi} \left ( \frac{3\vect{m} \cdot \vect{r}}{|\vect{r}|^5}  \vect{r} - %\frac{1}{|\vect{r}|^3} \vect{m} \right ) \ .
%\end{equation}
The entire computational domain can be discretized into numerous elements and each element can be approximated by one dipole at the center when the element size is sufficiently smaller than the distance between the magnet and the plasma.
Suppose the total number of discrete dipoles is $D$, then the total magnetic field is a summation over the dipoles,
\begin{equation} \label{eq:dipole}
    \vect{B}_{M} = \frac{\mu_0}{4\pi} \sum_{i=1}^{D} \left (\frac{3 \vect{m}_i \cdot \vect{r}_i}{|\vect{r}_i|^5} \vect{r}_i - \frac{1}{|\vect{r}_i|^3} \vect{m}_i \right) \ ,
\end{equation}
where $\ds \vect{m}_i = \sum_{j=x,y,z} m_i^j \vect{e}_j$ is the magnetic moment of $i$-th magnetic dipole,  $\ds \vect{r}_i = \sum_{j=x,y,z} r_i^j \vect{e}_j$ the position vector from the source to the evaluation point.

\subsection{Constraint on the maximum magnetization}
One of the explicit constraints that we have to follow is that the magnetization cannot exceed the remanence of materials ($B_r$).
The remanence varies with materials and material grades.
For example, neodymium-iron-boron(NdFeB) magnet, which is one of the most common magnets, normally has a remanence of 1.0 $\sim$ 1.5 T, while $\mathrm{Fe}_{16}\mathrm{N}_2$ magnets may have $B_r=2.9$ T \cite{Wang2020}.
In this paper, we will choose $B_r = 1.4$T for all the calculations, as this is one of the most widely commercially available magnets on the market.
However, it doesn't mean the remanence is fixed.
One could change the value of $B_r$ when using different materials.

The magnetic moment of each element is the volume integral of the magnetization, $\vect{m} = \iiint \vect{M} \dd{V}$.
If we consider homogeneous magnetization and take the relative permeability as $\mu=1$, each dipole will have a maximum allowable magnetic moment, which is then calculated as ${m_0}_i= B_r/{\mu_0} V_i$ ($V_i$ is the volume for the $i$-th element and $\mu_0$ the vacuum permeability).
% which is almost true with the present materials,
To better represent the constraint of the maximum magnetization, we shall split the magnetic moment of each dipole into magnitude and orientation.
A local spherical coordinates is used to represent the (unit) orientation,
\begin{equation}
    \vect{m}_i(\theta_i, \phi_i) = {m_0}_i \{\sin{\theta_i} \cos{\phi_i} \vect{e}_x + \sin{\theta_i} \sin{\phi_i} \vect{e}_y + \cos{\theta_i} \vect{e}_z \} \ .
\end{equation}
$\theta_i \in [-\pi, \pi]$ is the inclination angle and $\phi_i \in [-\pi, \pi]$ the azimuth angle.
Instead of using three components in Cartesian coordinates, we reduce the number of variables to two (\{$\t$, $\phi$\}) and explicitly enforce the constraint on the maximum magnetization ($m_0$).

\subsection{The density method}
The density method in topology optimization, which is also called \textit{Solid Isotropic Microstructure with Penalization} (SIMP) \cite{Rozvany1991}, represents the presence of material in each element with a normalized density varing from 0 to 1,
\begin{equation}
    \rho = 
    \begin{cases}
    \text{0 : no material;} \\
    \text{1 : material.}
    \end{cases}
\end{equation}
%The micro-structure inside each element is not considered and the material is assumed to be homogeneous.
In numerical optimizations, it is generally more complicated to do discrete optimization than optimizing continuous parameters.
Therefore, the density is allowed to be continuous.
For the consideration of simplifying assembly and reducing purchase cost, one would expect the magnets have uniform strength (and size).
In other words, the intermediate values of the normalized density ($0 < \rho <1 $) should be avoided.
A special technique is applied to penalize the intermediate values.
The normalized density is written in an exponential form, $\rho = p^q$, where $p\in[0,1]$ is the free parameter and $q$ the penalize coefficient.
The penalize coefficient is specified by users and the larger $q$ is set, the more stiff the distribution is.
By using a relatively large value of $q$, like $q=7$, one could achieve a polarized distribution for the normalized density. 
Now, the final representation for the magnetic moment of each dipole becomes 
\begin{equation}
    \vect{m}_i(p_i, \theta_i, \phi_i) = {p_i}^q \, {m_0}_i\{\sin{\theta_i} \cos{\phi_i} \vect{e}_x + \sin{\theta_i} \sin{\phi_i} \vect{e}_y + \cos{\theta_i} \vect{e}_z \} \ ,
\end{equation}
where the free variables for the $i$-th dipole are $p_i$, $\theta_i$, and $\phi_i$.
$p_i$ controls the density and $\{\theta_i, \phi_i \}$ determines the orientation.
${m_0}_i$ comes from the product of the maximum allowable magnetization and the element volume.

\subsection{Objective functions}
%In general stellarator optimization problems, MHD equilibrium with desired properties is obtained first and then the coils are designed by a ``reverse-engineering" process.
To produce the required magnetic field, the primary objective function in coil optimization codes is usually the Neumann boundary condition, which is the normal magnetic field on a prescribed toroidal surface $\cal S$.
Likewise, to optimize the permanent magnets, we can employ an objective function for the magnetic field in the following form,
\begin{equation} \label{eq:fb}
    F_B(p, \theta, \phi) = \iint_{\cal S} \left ( \vect{B}_{M} \cdot \vect{n} - B_n^{tgt} \right )^2 \dd{a} \ ,
\end{equation}
where $B_n^{tgt}$ is the target normal field.
The target normal field is one of the inputs in this problem.
Normally, the last-closed-flux-surface (LCFS) is used as the plasma boundary.
Then the target normal field is the inverse of the normal components from plasma currents and coils, as the overall magnetic field has no normal components on the plasma surface.
The normal field at an arbitrary point on the plasma boundary generated by the dipoles can be calculated as,
\begin{equation}
    \vect{B}_{M} \cdot \vect{n} (\t_s, \z_s) = \frac{1}{|\vect{N}|} \sum_{i=1}^{D} \vect{g}_i \cdot \vect{m}_i \ , 
\end{equation}
where we parameterize the plasma boundary using the poloildal angle $\t_s$ and toroidal angle $\z_s$. 
$\vect{N}$ is the normal vector of the plasma boundary ($\vect{n}$ the unit normal vector) and $\vect{g}$ is the ``inductance'' matrix.
Since the geometric quantities do not change during the optimizations, we can compute $\vect{g}$  once and then store it in the memory,
\begin{equation} \label{eq:inductance}
\ds \vect{g}_i (\t_s, \z_s) = \frac{\mu_0}{4\pi} \left (\frac{3 \vect{r}_i \cdot \vect{N}}{|\vect{r}_i|^5} \vect{r}_i - \frac{1}{|\vect{r}_i|^3} \vect{N}  \right) \ .
\end{equation}
The first-order derivatives of $F_B$ can be analytically computed,
\begin{align} 
    \label{eq:derivatives01} \pdv{F_B}{p_i} & = 2 \int_{\t_s} \int_{\z_s} \left ( \frac{1}{|\vect{N}|} \sum_{j=1}^{D} \vect{g}_j \cdot \vect{m}_j - B_n^{tgt} \right ) q {p_i}^{q-1} \, {m_0}_i\left ( g_i^x \sin{\theta_i} \cos{\phi_i} + g_i^y \sin{\theta_i} \sin{\phi_i} + g_i^z \cos{\theta_i} \right ) \dd{\t_s} \dd{\z_s} \ ,  \\ 
    \label{eq:derivatives02} \pdv{F_B}{\t_i} & = 2 \int_{\t_s} \int_{\z_s} \left ( \frac{1}{|\vect{N}|} \sum_{j=1}^{D} \vect{g}_j \cdot \vect{m}_j - B_n^{tgt} \right )  {p_i}^{q} \, {m_0}_i\left ( g_i^x \cos{\theta_i} \cos{\phi_i} + g_i^y \cos{\theta_i} \sin{\phi_i} - g_i^z \sin{\theta_i} \right ) \dd{\t_s} \dd{\z_s} \ ,   \\ 
    \pdv{F_B}{\phi_i} & = 2 \int_{\t_s} \int_{\z_s} \left ( \frac{1}{|\vect{N}|} \sum_{j=1}^{D} \vect{g}_j \cdot \vect{m}_j - B_n^{tgt} \right ) {p_i}^{q} \, {m_0}_i\left ( - g_i^x \sin{\theta_i} \sin{\phi_i} + g_i^y \sin{\theta_i} \cos{\phi_i} \right ) \dd{\t_s} \dd{\z_s} \ . \label{eq:derivatives03}
\end{align}
%In the above equations, the surface integrals are performed on the plasma boundary.

The inverse problem of the Biot-Savart law is ill-posed and coil optimization codes usually have other penalty functions on coil geometry to obtain engineering favorable solutions.
For the permanent magnet problem, we adopt a regularization/penalization term on the total number of magnetic moments, which is calculated as,
\begin{equation} \label{eq:fm}
    F_M(p) = \sum_{i=1}^{D} |\vect{m}_i|^2 \ .
\end{equation}
$F_M$ is determined by the density parameter only.
The derivative of $F_M$ with respect to $p_i$ is
\begin{equation} \label{eq:derivatives04}
    \pdv{F_M}{p_i} = 2q {p_i}^{2q-1} \left ( {m_0}_i\right)^2 \ , 
\end{equation}
while $\partial F_M / \partial \theta_i$ and $\partial F_M / \partial \phi_i$ are zero.

There is another constraint that we have to consider for a practical design.
The access to the plasma is essential for heating and diagnostic systems, which implies the magnets cannot cover the entire space.
Thus, there are forbidden regions where the magnets cannot be present.
Instead of employing new objective functions to preserve the forbidden regions, we have a more straight-forward way.
For the elements inside the forbidden regions, the associated parameters $(p_i, \theta_i, \phi_i)$ will not be varied and $p_i$ is zero during the optimization.
By doing so, we can exclude the magnets inside the forbidden regions.

The final objective function is then the weighted summation of $F_B$ and $F_M$,
\begin{equation} \label{eq:target}
    F(p, \theta, \phi) = F_B + \lambda F_M \ ,
\end{equation}
where $\lambda$ is a user-specified weight for the penalization.

\subsection{Numerical implementation}
The FAMUS (Flexbile Advanced Magnets Used for Stellarators) code has been developed, on the foundation of the FOCUS \cite{FOCUS00} code.
The problem of finding optimal magnet layout can be expressed in mathematical languages as,
\begin{align}
%    \argmin_{p_i,\theta_i,\phi_i} {\iint_S \left [ \vect{B}_{M}(p_i,\theta_i,\phi_i) \cdot \vect{n} - B_n^{tgt} \right ]^2 \dd{a} + \lambda \sum_{i=1}^{D} |\vect{m}_i (p_i,\theta_i,\phi_i)|^2 } \nonumber \\ 
    \text{min} & \quad   F(p_i,\theta_i,\phi_i) = {\iint_S \left [ \vect{B}_{M}(p_i,\theta_i,\phi_i) \cdot \vect{n} - B_n^{tgt} \right ]^2 \dd{a} + \lambda \sum_{i=1}^{D} |\vect{m}_i (p_i,\theta_i,\phi_i)|^2 } \\
    \text{s. t.}  & \quad p_i\in[0,1], \; \theta_i \in [-\pi, \pi], \; \phi_i \in [-\pi, \pi], \; i=1, \cdots, D \ .
\end{align}
The variables are the density and orientation of each magnetic dipole, and they have explicit bounds.
For simplification, we assume the magnetization doesn't change when the background field changes.
In other words, there is no interaction between each dipole and thus the variables are separable.
It is beneficial to use message passing interface (MPI) for parallelization and each CPU can handle a sub-group of dipoles.
To minimize the target function, we use a limited-memory quasi-Newton algorithm (the L-BFGS-B library) \cite{lbfgsb} with defined bounds.
The gradient is calculated analytically using the formulas in \Eqn{derivatives01}, (\ref{eq:derivatives02}), (\ref{eq:derivatives03}) and (\ref{eq:derivatives04}).
To reduce the number of degrees of freedom, FAMUS can enforce toroidal periodicity and/or stellarator symmetry \cite{stellarator_symmetry} if necessary.

To retain the precision of modeling, FAMUS tends to use a large number of dipoles, which means the dimensionality is (extremely) large.
It is not an issue for the computation speed since the efficiency of MPI is considerably high and the usage of analytically calculated derivatives makes sure the problem can be easily scaled-up.
However, the large number of variables makes it easy to get trapped in local minima.
An example of FAMUS finding different solutions with different initial guesses is shown in Appendix \ref{local_minimum}.
A good initial guess is thus important for successful optimizations.
In Ref. \cite{ZhuPM01}, we introduced a linear method to design perpendicular magnets using current potential.
It could provide a good initial guess for FAMUS.
Besides, we can linearly solve the magnitude and/or orientation of each magnetic dipole if we don't enforce the constraint over the maximum magnetization, as shown in Appendix \ref{least_square}.
We might not be able to incorporate the constraint of the maximum magnetization without further modifications, but the linear solved solution can be truncated manually and should provide a relative good starting point for FAMUS nonlinear optimizations.
In addition, other approaches, like the one proposed by Helander \etal \cite{HelanderPM} and REGCOIL\_PM \cite{REGCOIL_PM}, can also be used for initialization.
%Suppose the number of toroidal periodicity is $N_p$ and the stellarator symmetry is enforced.
%A straightforward implementation is to rotate the dipoles along the z-axis (for the toroidal periodicity) and the x-axis (for the stellarator symmetry).
%The total number of dipoles is now $D \times 2 \times N_p$, where we assume the number of unique dipoles is $D$.
%Then one can calculate the magnetic field for all the dipoles.
%When the number of unique dipoles, $D$, is large, storing all the dipoles may use a huge amount of memory.
%Instead, we adopt a different strategy in FAMUS.

\section{Benchmark with the surface magnetization method} \label{benchmark}
%The surface magnetization method reveals the connection between current potential and surface magnetization.
To validate the FAMUS code, we can compare the solution optimized by FAMUS with the one obtained by the surface magnetization method \cite{ZhuPM01}, which shall be called as ``the linear method'' hereafter.
For this benchmark, a two-period rotating elliptical stellarator was used.
Details of the numerical setup for the linear method can be found in Sec. 3.1 of the Ref. \cite{ZhuPM01}.
%The boundary shape was described in Ref. \cite{ZhuPM1} and a central current was used to provide the toroidal field.
By using the linear method, a single-layer of magnetic dipoles ($128\times 128$ per period) that are perpendicular to the winding surface was obtained to support the vacuum configuration together with a toroidal field.
%The magnitude of the magnetic moment of the dipoles has helical patterns spatially.

\begin{figure}[htb]
    \centering 
    \includegraphics[width=0.8\textwidth]{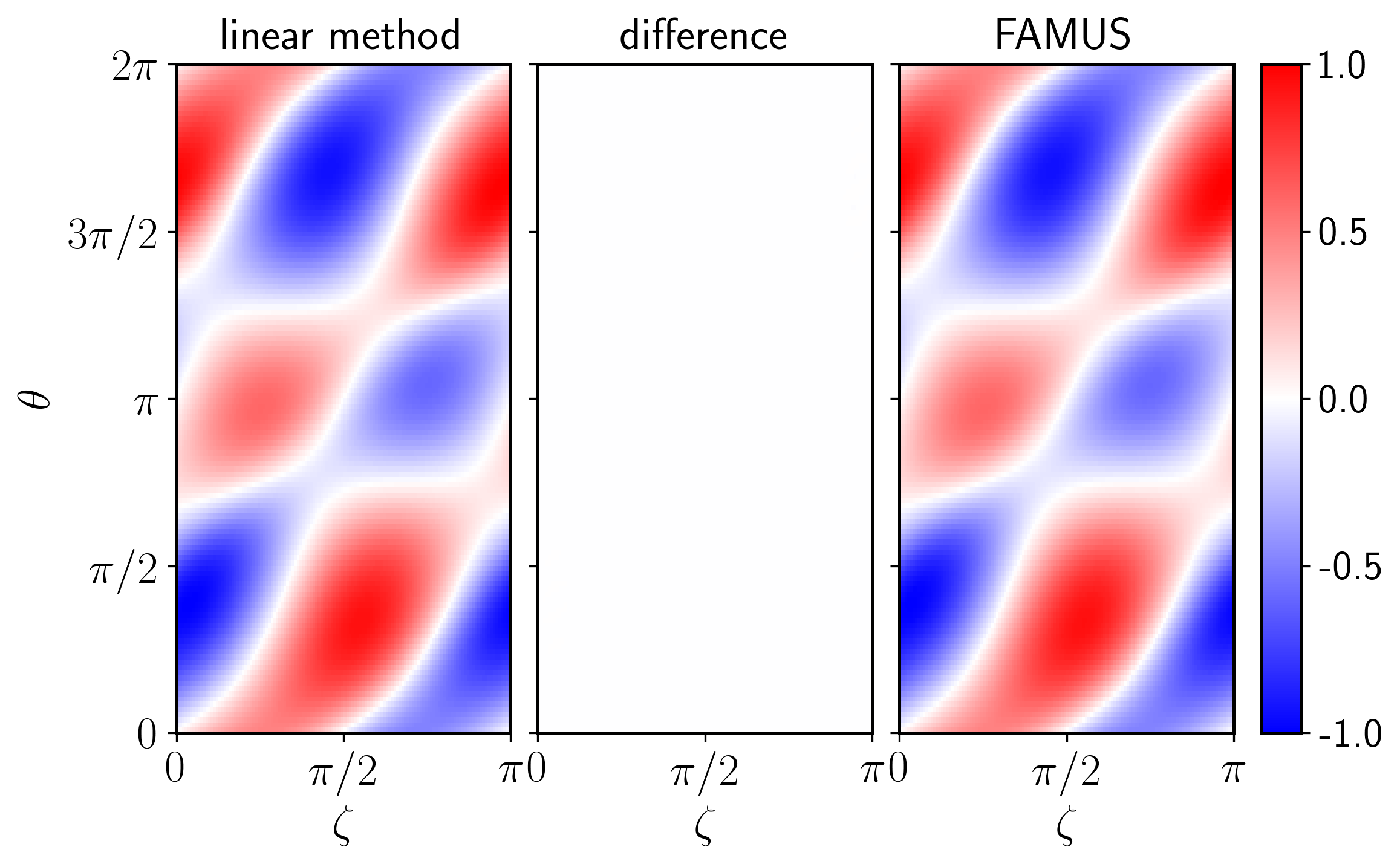}
    \caption{Distributions of the normalized density for the single-layer dipoles from the linear method (left) and FAMUS (right). The difference between the two solutions (middle) is almost zero.}
    \label{fig:surf_mag_comp}
\end{figure}

To prepare FAMUS runs, the positions of the magnetic dipoles were input to FAMUS and their orientations were fixed to be perpendicular to the winding surface (pointing outwards).
The allowable moment ${m_0}_i$ for each magnetic dipole was set to the maximum value of the magnetic moment in the linear method solution.
Therefore, the normalized density in the linear method solution is in the range of $[-1, 1]$ and the distribution is shown in \Fig{surf_mag_comp}.
In FAMUS runs, only the normalized density of each dipole was allowed to vary in the same range of $p_i \in [-1, 1]$.
The number of free parameters was 16384. 
The initial density of each dipole was set to a small, non-zero number, $1.0\tento{-3}$ and the exponential factor was $q=1$.
%As the magnets also point inwards, we set the bounds of the density function to be $p \in [-1, 1]$ to allow reverse directions.
$F_B$ was reduced from $8.73\tento{-3}$ to $5.20\tento{-13} \ \mathrm{T}^2 \mathrm{m}^2$ in 100 iterations.
The elapsed wall time was 389.28 seconds with 32 CPUs (Intel Xeon E5-2670@2.60GHz).
%The final normal field error can be as low as $1.04\tento{-14}$.
\Fig{surf_mag_comp} shows the comparison between the distribution of the density optimized by FAMUS and the target solution from the linear method.
The difference between the two solutions are negligible and the average difference in the normalized density is  $9.94\tento{-4}$ (maximum difference $8.91\tento{-3}$).
The results indicate that FAMUS can find exactly the same solution (to some numerical tolerance) as the linear method if we restrict the magnets to be perpendicular only and relax the constraint on the maximum allowable magnetization.

\section{Applications on NCSX} \label{application}
Now we can apply FAMUS to practical configurations.
%Without loss of generality
For demonstration, we choose the quasi-axisymmetric stellarator, NCSX \cite{NCSX}, as the target configuration.
NCSX was partly built at the Princeton Plasma Physics Laboratory (PPPL) and unfortunately canceled due to cost overrun.
It was originally designed with 18 modular coils to generate the main magnetic field.
In our study, we are going to only use the 18 planar TF coils to provide the essential toroidal field.
One of the designed equilibria, C09R00, was scaled to have a volume-averaged magnetic field of $\langle B \rangle=0.5$ T, which is the maximum field produced by the TF coils.
Some other parameters of C09R00 are $\mathrm{N_{fp}} = 3$, $R_0=1.44$ m,
$a=0.32$ m, $V_{\mathrm{plasma}}=2.96 \mathrm{m^3}$, and $\langle \beta \rangle=4.09\%$.

To further embrace engineering constraints, we will re-use the NCSX vacuum vessel \cite{NCSX_vessel}, which has been built as well.
Ports have been opened for diagnostics and heating systems.
These ports are the explicit forbidden regions wherein permanent magnets can not be placed.
In this paper, we are going to consider the three largest ports with extensions, the vertical ports (upper and lower), the neutral-beam port and the port 4, while other ports can also be reserved if necessary.
% As shown in \Fig{ncsx_vessel}, the vessel and its accessories (ports, extensions) define a clear design space for permanent magnets: the magnets should be outside the vessel and have to stay away from the ports.
% 
% \begin{figure}
%     \centering
%     \includegraphics[width=\textwidth]{NCSX_vessel_coils.png}
%     \caption{NCSX vessel, ports and TF coils. Only two third of the vessel and TF coils are shown here. Three largest ports, the vertical ports (upper and lower), the neutral-beam port and the port 4, are shown with extensions.}
%     \label{fig:ncsx_vessel}
% \end{figure}

In the following, we will present three applications to the NCSX configuration. 
The first one fixes the magnet orientations and only the normalized density is varied.
The second one has arbitrary orientations to improve the magnetic field efficiency and uses less magnets.
The third one is incorporated with simplified geometries for the magnets.

\subsection{Permanent magnets in perpendicular direction} \label{perp_only}
We can now use FAMUS to improve the finite-thickness solution obtained by the linear method \cite{ZhuPM01}.
In that design, all the magnets are perpendicular to the vessel, pointing either inwards or outwards.
As we have discussed, it was not an overall optimization and did not explicitly incorporate the ports.

\begin{figure}[hbt!]
\centering
\begin{minipage}{.48\linewidth}
  \centering
  \includegraphics[width=\textwidth]{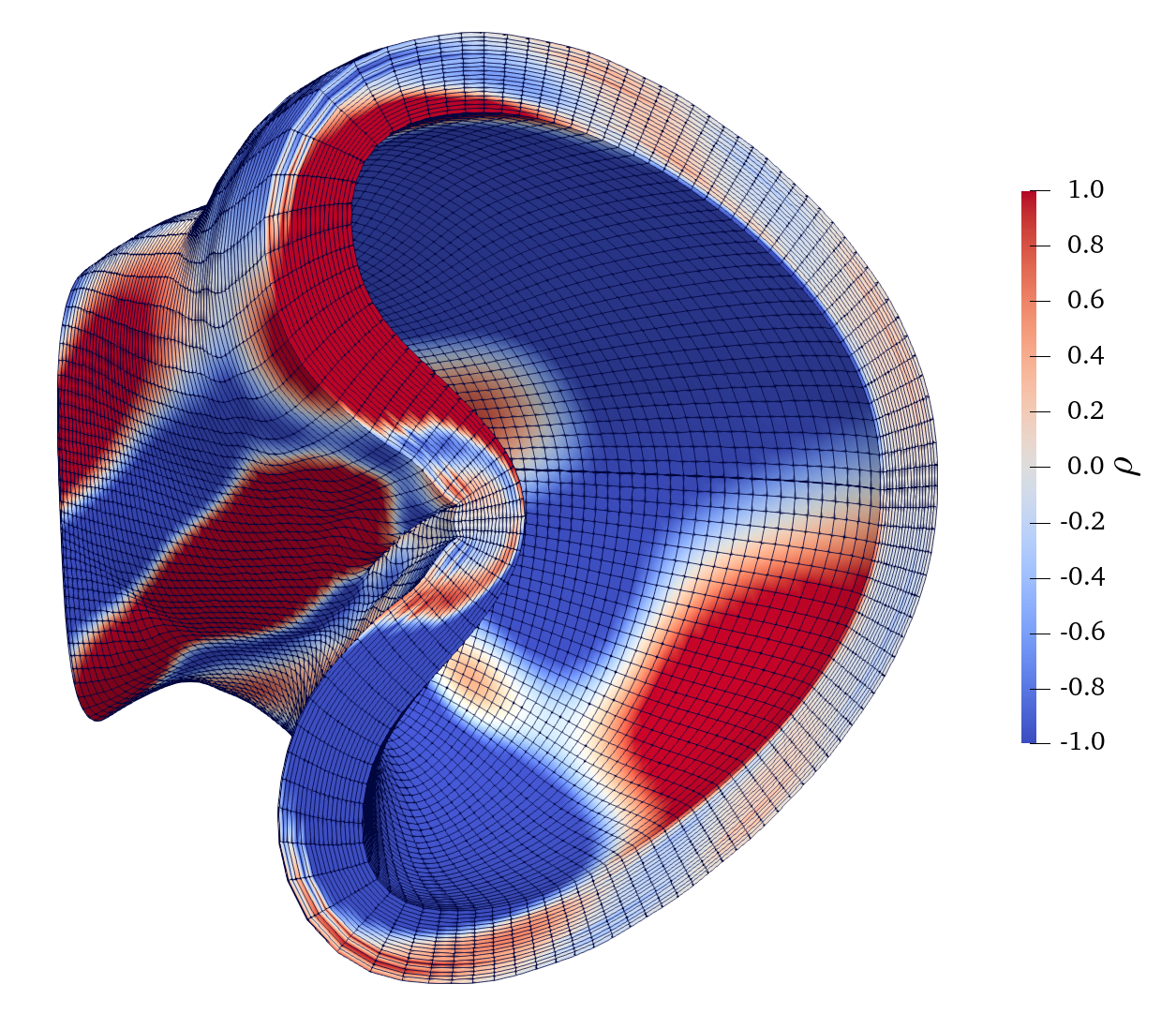}
  \captionof{figure}{The distribution of the normalized density for the multi-layer perpendicular magnetization from the linear method.}
  \label{fig:perp_initial}
\end{minipage}\hfill
\begin{minipage}{.48\linewidth}
  \centering
  \includegraphics[width=\textwidth]{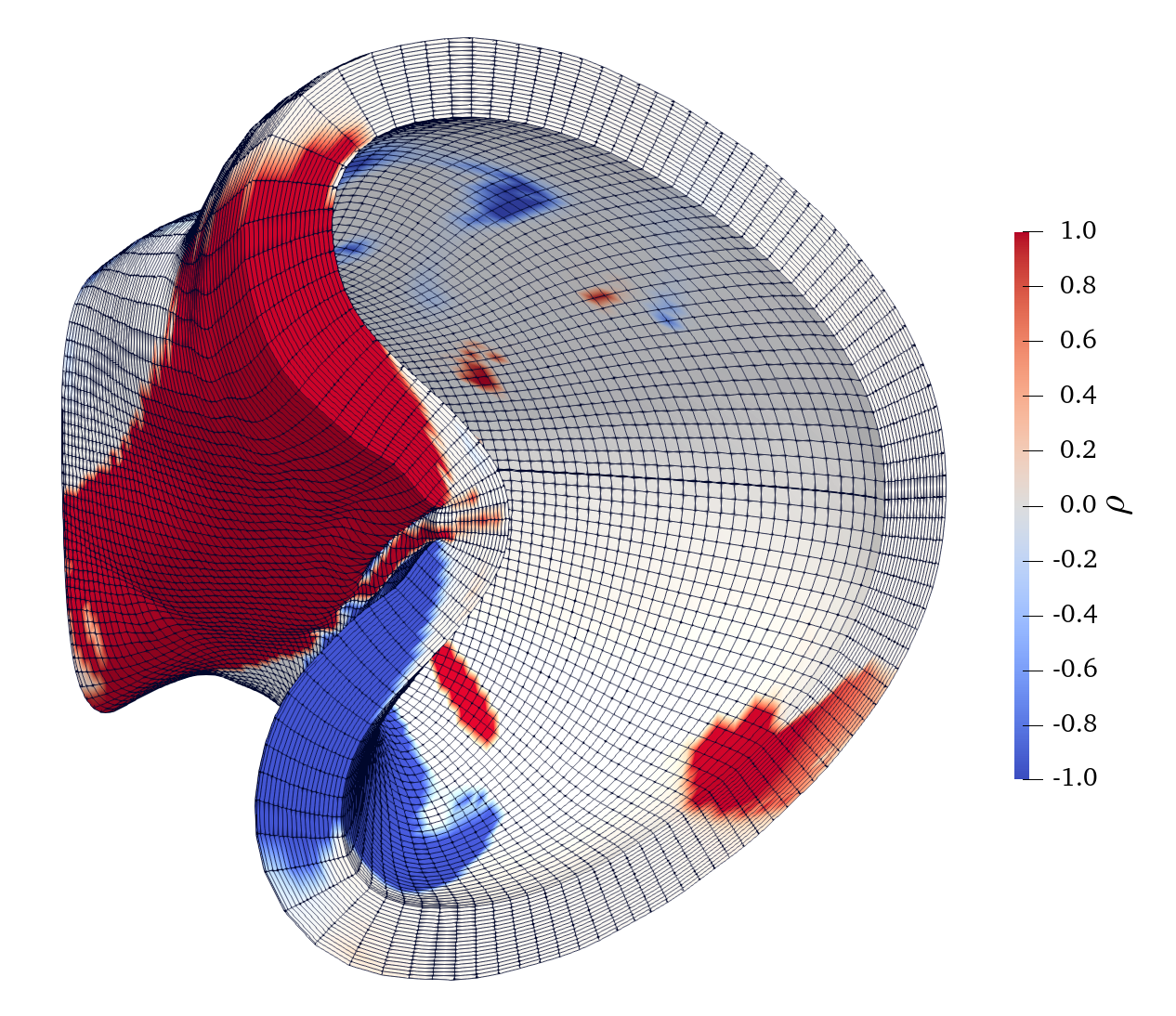}
  \captionof{figure}{The distribution of the normalized density for the FAMUS-optimized solution with fixed orientations.}
  \label{fig:perp_final}
\end{minipage}
\end{figure}

The magnetic dipoles from the linear method are lying on a sequence of surfaces offset from the vessel.
The innermost surface is 2 cm away from the vessel and the outermost is 22 cm away, while the distance between two adjacent surfaces is 1 cm.
The resolution of magnetic dipoles is {20 (radial) $\times 128$ (poloidal)  $\times 64$ (toroidal)} per half period.
The orientation of each dipole is aligned to be perpendicular to the supporting surface which they are lying on and we chose the outwards direction as the positive direction.
The density of each dipole is normalized to the allowable magnetic moment for each element, which is calculated as ${m_0}_i = B_r/\mu_0 V_i$ ($B_r=1.4$ T,  $V_i$ the volume of each element).
\Fig{perp_initial} shows the distribution of the normalized density for the multi-layer linear solution.
The normal field error, $F_B$, as calculated in \Eqn{fb}, is $3.99\tento{-4} \mathrm{T}^2 \mathrm{m}^2$ per half-period (same definition is used for all the following calculations).
The total amount of magnetic moment, $\sum_{i=1}^{D}  |m_i|$, is $7.53\tento{5} \mathrm{A} \cdot \mathrm{m}^2$ per half-period, which is equivalent to the magnetic moment from the NeFeB magnet with a volume of $0.68 \ \mathrm{m}^3$.

The linear method solution is used as the initial guess for FAMUS, while the magnetic dipoles in the forbidden regions are filtered out (fixed to be zero during the optimization).
To penalize the intermediate values in the density function, we chose the penalization coefficient $q=7$.
Free variables were the normalized density of each magnetic dipole, $p_i \in [-1, 1]$.
The orientation was fixed during the optimizations.
The total number of variables was 154843 with 8997 elements inside the forbidden regions.
FAMUS can further minimize the normal field, optimize the layout, reduce the magnet volume.
However, we found that initializing with $p_i=1.0$ yielded a better solution with larger access on the outboard side.
\Fig{perp_final} shows the distribution of the normalized density optimized from the initial guess of $p_i=1.0$ with $q=7$.
In 200 iterations (26 minutes of wall time with 256 CPUs), $F_B$ was reduced to $2.91\tento{-5} \mathrm{T}^2 \mathrm{m}^2$ and the total used magnetic moment was $5.49\tento{5} \  \mathrm{A} \cdot \mathrm{m}^2$ (equivalent to $0.50 \  \mathrm{m}^3$ NeFeB magnet).

\begin{figure}[hbt!]
    \centering
    \includegraphics[width=0.8\linewidth]{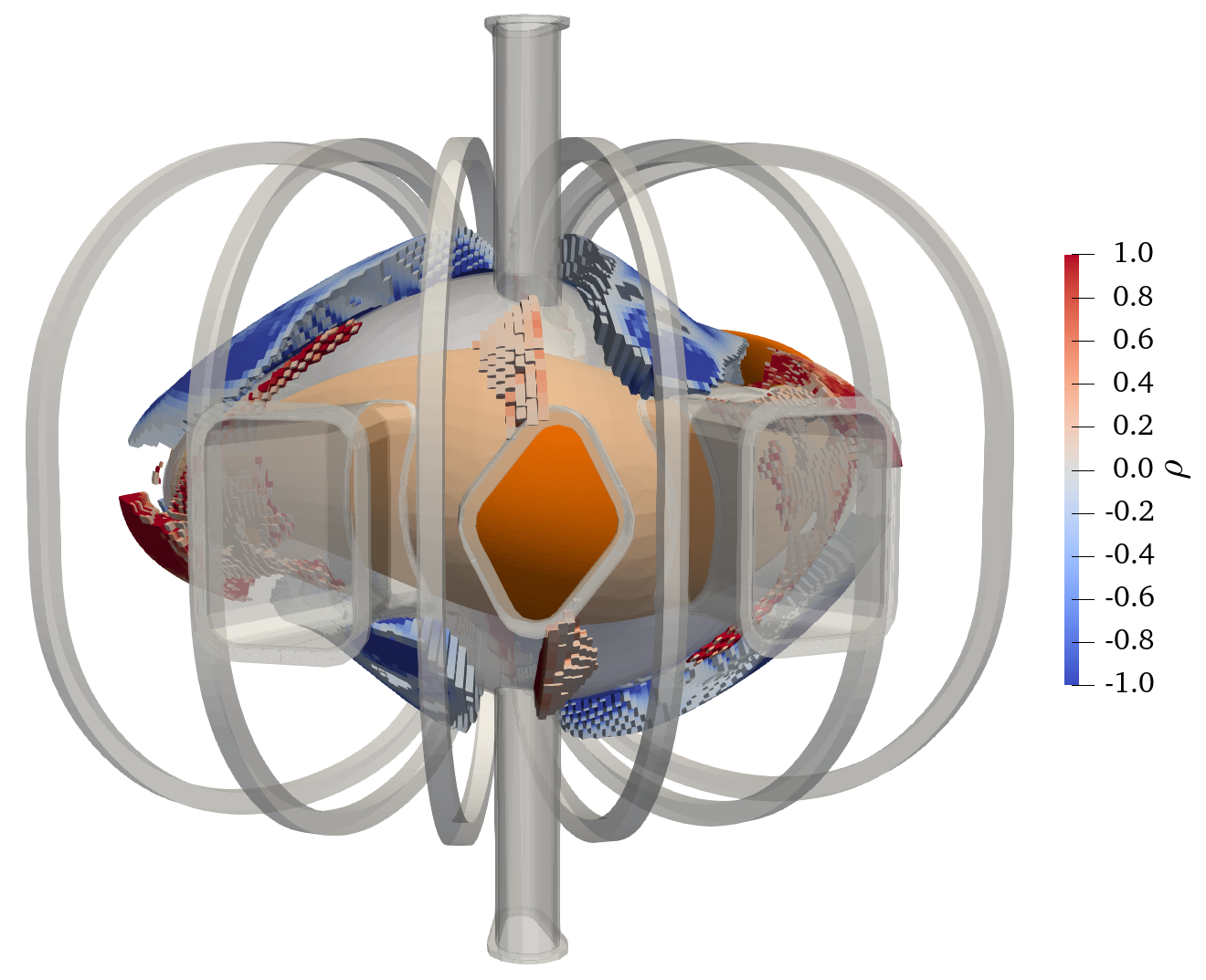}
    \caption{Outboard view of the perpendicular permanent magnets together with TF coils for the half-Tesla NCSX configuration. Magnets with normalized density $<0.1$ are removed for clarity.}
    \label{fig:normal_outboard}
\end{figure}

\begin{figure}[hbt!]
\centering
\begin{minipage}{.48\linewidth}
    \centering
    \includegraphics[width=\linewidth]{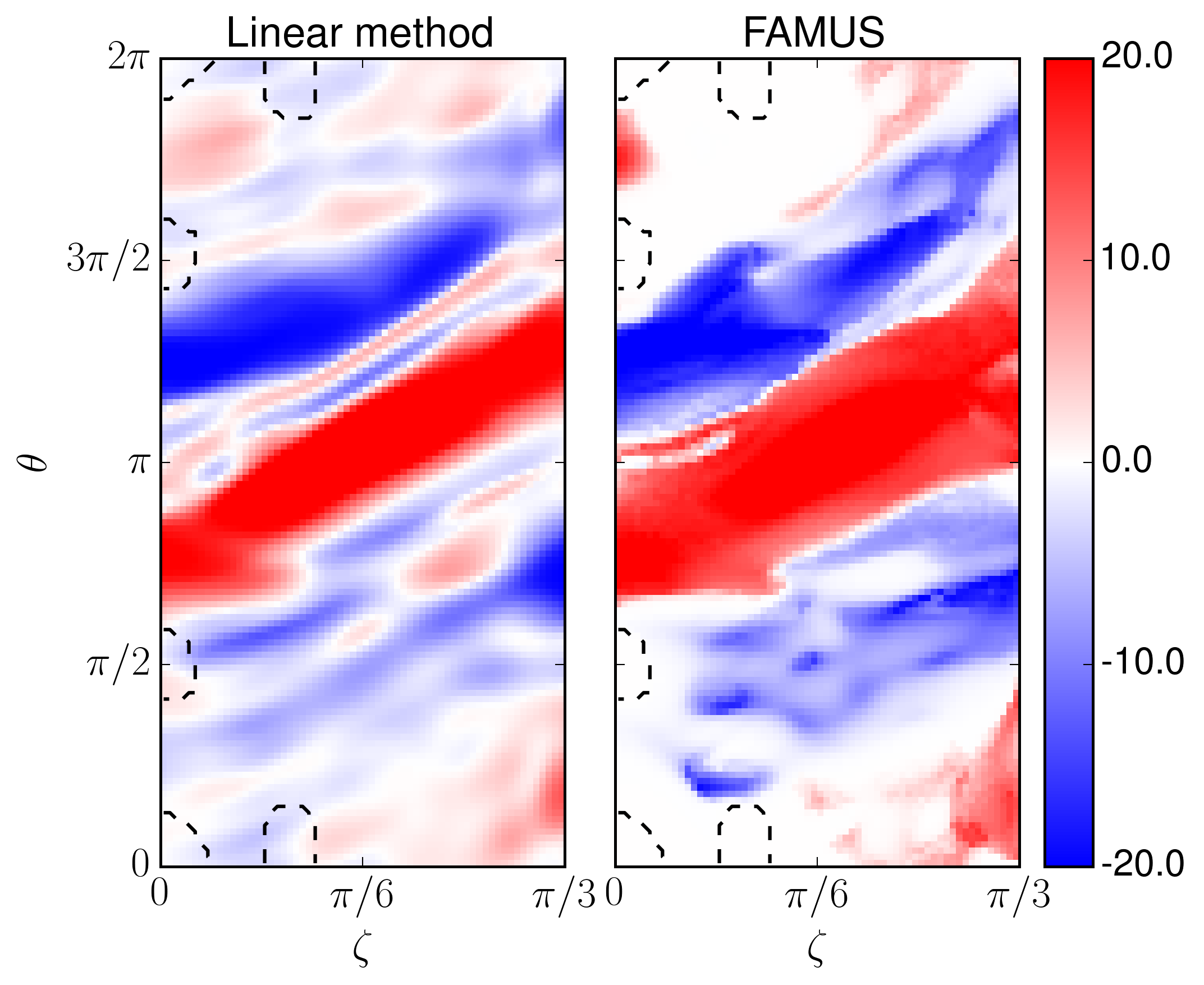}
    \caption{Comparison of the radial accumulation of the normalized density on the $\t - \z$ plane for the solutions from the linear method and FAMUS. Dashed lines represent the intersections of the reserved ports and the outermost layer of meshes.}
    \label{fig:density_comp}
\end{minipage}\hfill
\begin{minipage}{.48\linewidth}
    \centering
    \includegraphics[width=\linewidth]{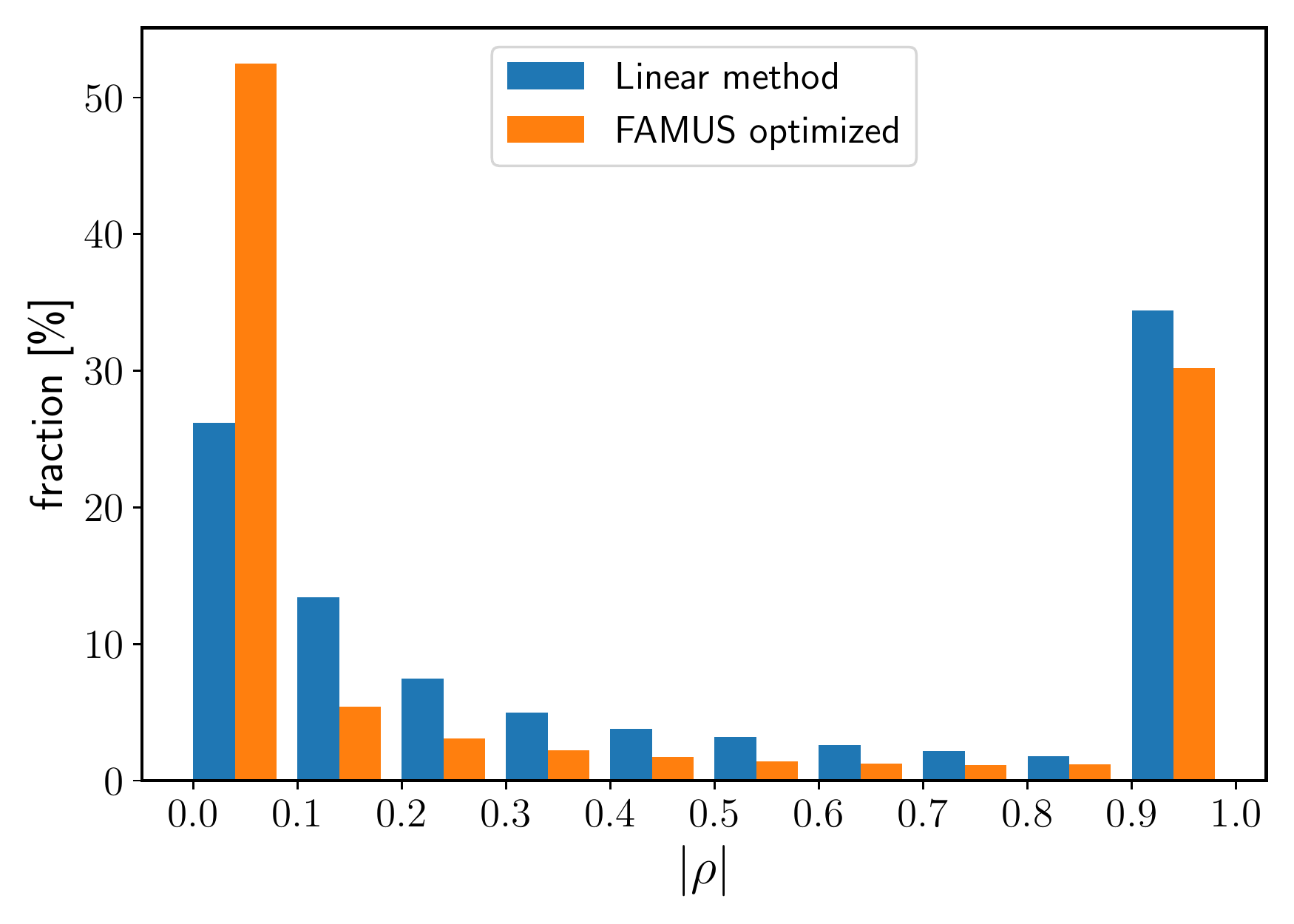}
    \caption{Histograms of the normalized densities for the linear method solution and FAMUS-optimized solution. The intervals are of equal size (0.1) in the range of [0.0, 1.0] for both cases. The rectangles are not plotted with full width to clarify the distinction.}
    \label{fig:rho_comp}
\end{minipage}
\end{figure}

Significant difference in the distribution of the normalized density can be observed when comparing \Fig{perp_initial} and \ref{fig:perp_final}.
Since we are using structured grids for the meshes of dipoles, we can compute the radial accumulation of the normalized density. 
As shown in \Fig{density_comp}, the original solution from the linear method covers almost the entire space.
The FAMUS optimized solution can explicitly incorporate ports and has large area of zero-density regions where no magnets are needed.
The magnets are mostly localized on the inboard side.
It can be more clearly observed when viewing from the outboard side, as illustrated in \Fig{normal_outboard}, in which we removed the magnets with $|\rho|<0.1$.
The almost-open outboard side can provide considerably large access to the plasma, which can be used for heating systems, diagnostics and maintenance.
The large access to the plasma would be exceptionally attractive to future fusion reactors which rely on remote maintenance.

We can also check if the density method is working as expected.
In \Fig{rho_comp}, we plot out the histograms of the normalized density for the linear method solution and FAMUS-optimized solution.
The FAMUS-optimized solution has a more polarized distribution.
More than half of the magnets (52.49\%) have a density smaller than 0.1 and 30.18 \% of the magnets have a density larger than 0.9.
The intermediate values are indeed penalized by adopting the exponential function of the density.
The accuracy will be mostly retained when the magnets present only at the places $\rho>0.9$.

\subsection{Permanent magnets with arbitrary orientations}
\begin{figure}[hbt!]
  \centering
  \includegraphics[width=0.5\linewidth]{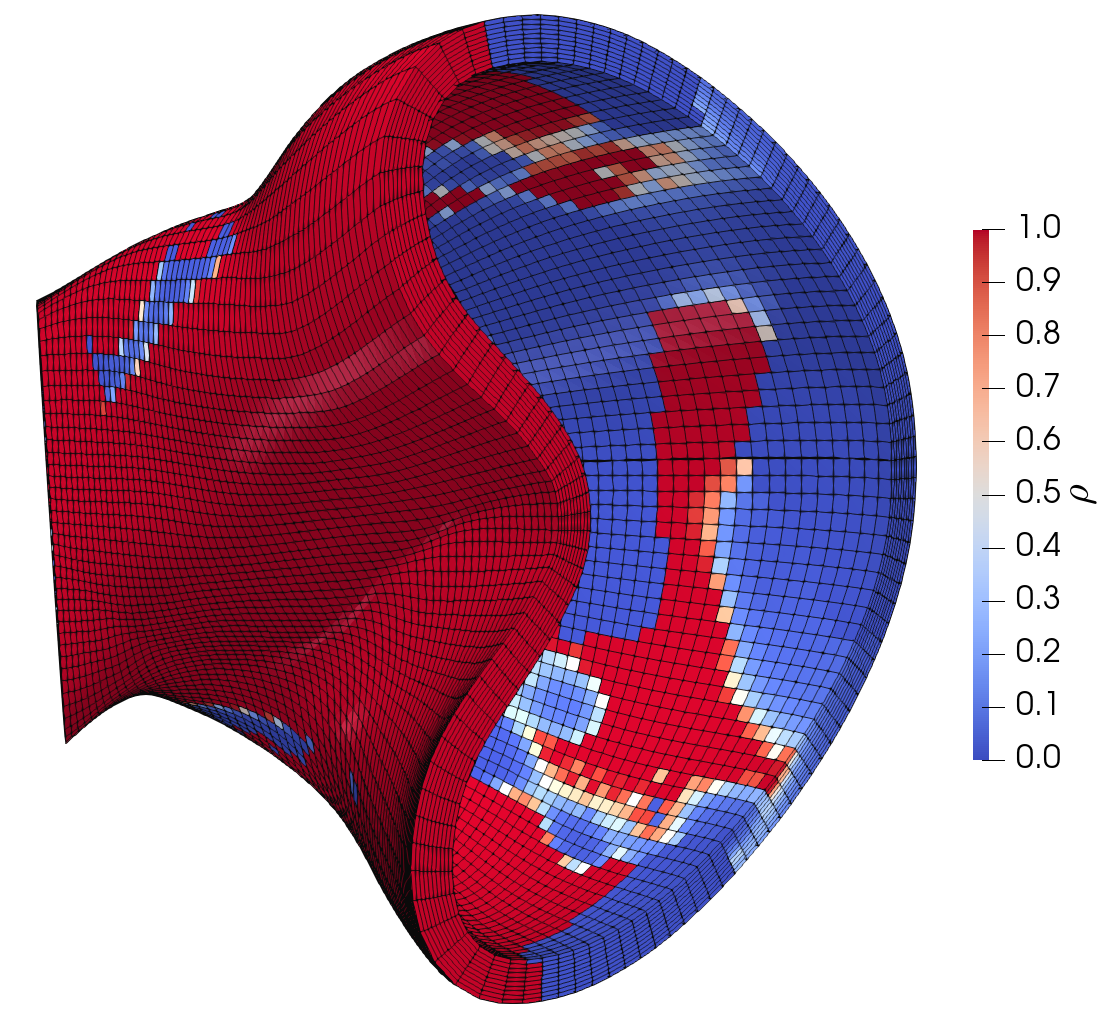}
  \captionof{figure}{Distribution of the normalized density for the 10-cm-thick orientation optimized solution.}
  \label{fig:orient_10cm_density}
\end{figure}

We can now relax the restriction of orientation to improve the magnetic efficiency.
The ``Halbach'' array \cite{Halbach1980} can improve the magnetic efficiency by a factor of two in simple 1D and 2D cases.
Thus, we can try to reduce the thickness of the design space from 20 cm to 10 cm.
The resolution of the structured meshes is now {10 (radial) $\times 128$ (poloidal)  $\times 64$ (toroidal)} for half period.
Again, stellarator symmetry is enforced.
%and the dipoles in the forbidden regions are fixed to have zero density.
To avoid being trapped in local minima, we initialized the dipole orientation to be perpendicular to the vessel and split the optimization into three steps.
First, the density of each dipole was fixed to be one (including those in the forbidden regions) and the orientations were optimized to only reduce the normal field error, $F_B$.
$F_B$ was decreased from $7.07 \tento{-2}$ to $7.71 \tento{-6}$ \ $\mathrm{T}^2 \mathrm{m}^2$ in 100 iterations.
Subsequently, the density was allowed to be varied and both $F_B$ and $F_M$ were minimized.
Also, $q=8$ was used to penalize the intermediate values for the density.
In 200 iterations, $F_B$ was further reduced to $2.81 \tento{-6} \mathrm{T}^2 \mathrm{m}^2$ while the total amount of the used magnetic moment was reduced by 31.1\%.
In the third step, the magnetic dipoles in the forbidden regions were filtered out.
The orientations and normalized densities of the other dipoles were furthur optimized.
The final solution is shown in \Fig{orient_10cm_density}, while $F_B = 6.57\tento{-6} \  \mathrm{T}^2 \mathrm{m}^2$ and the total used magnetic moment is $4.75 \tento{5} \ \mathrm{A} \cdot \mathrm{m}^2$ (equivalent to $0.43 \  \mathrm{m}^3$ NeFeB magnet)

\begin{figure}[hbt!]
  \centering
  \includegraphics[width=0.8\linewidth]{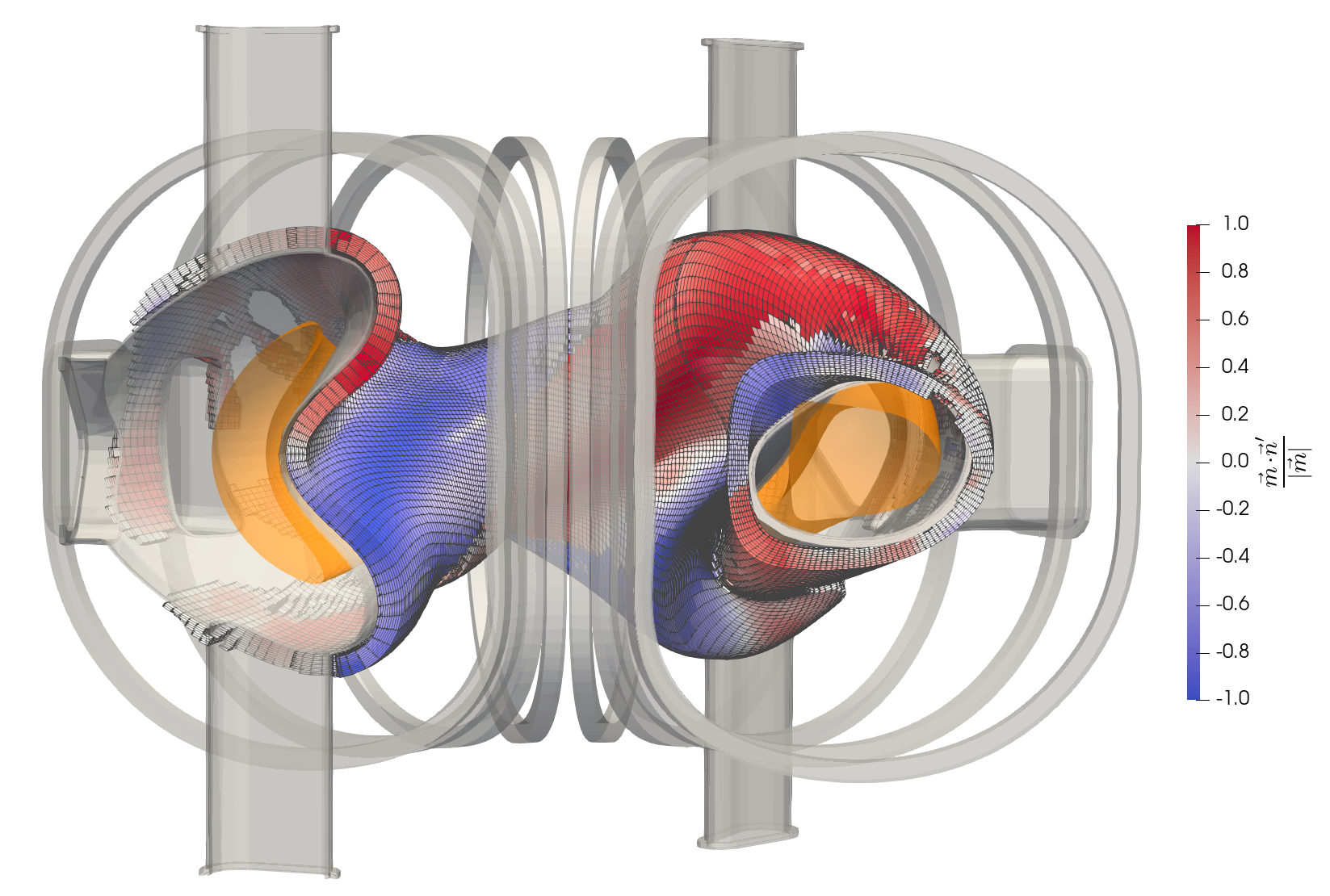}
  \captionof{figure}{Orientations of the magnetic dipoles. The colors stand for the cosine value of the angle between the dipole orientation and the vessel normal. The dipoles with $|\rho|<0.1$ have been removed.}
  \label{fig:oreint_10cm_mdotn}
\end{figure}

We can check the orientation of each magnetic dipole.
In \Fig{oreint_10cm_mdotn}, we plot the value of ${\vect{m} \cdot \vect{n}'}/{|\vect{m}|}$ where $\vect{n}'$ is the vessel normal.
This quantity represents the angle between the dipole orientation and the vessel normal.
The dipoles with $|\rho|<0.1$ have been clipped and most the dipoles are located on the inboard side as well.
On the inboard side, tilt stripes are observed.
At the bean-shaped cross-section, the dipoles are pointing outwards at the upper half, and inwards at the lower regions.
Between the two regions, the dipoles are mostly tangential to the vessel normal.
This is somehow similar to a Halbach array with the pattern of $\uparrow \rightarrow \downarrow \leftarrow \uparrow$.
Of course, the geometries are fully 3-D and  the actual distribution is much more complicated than ideal Hablbach arrays.

\subsection{Permanent magnets with simplified geometry}

\begin{figure}[hbt!]
    \centering
    \includegraphics[width=0.6\linewidth]{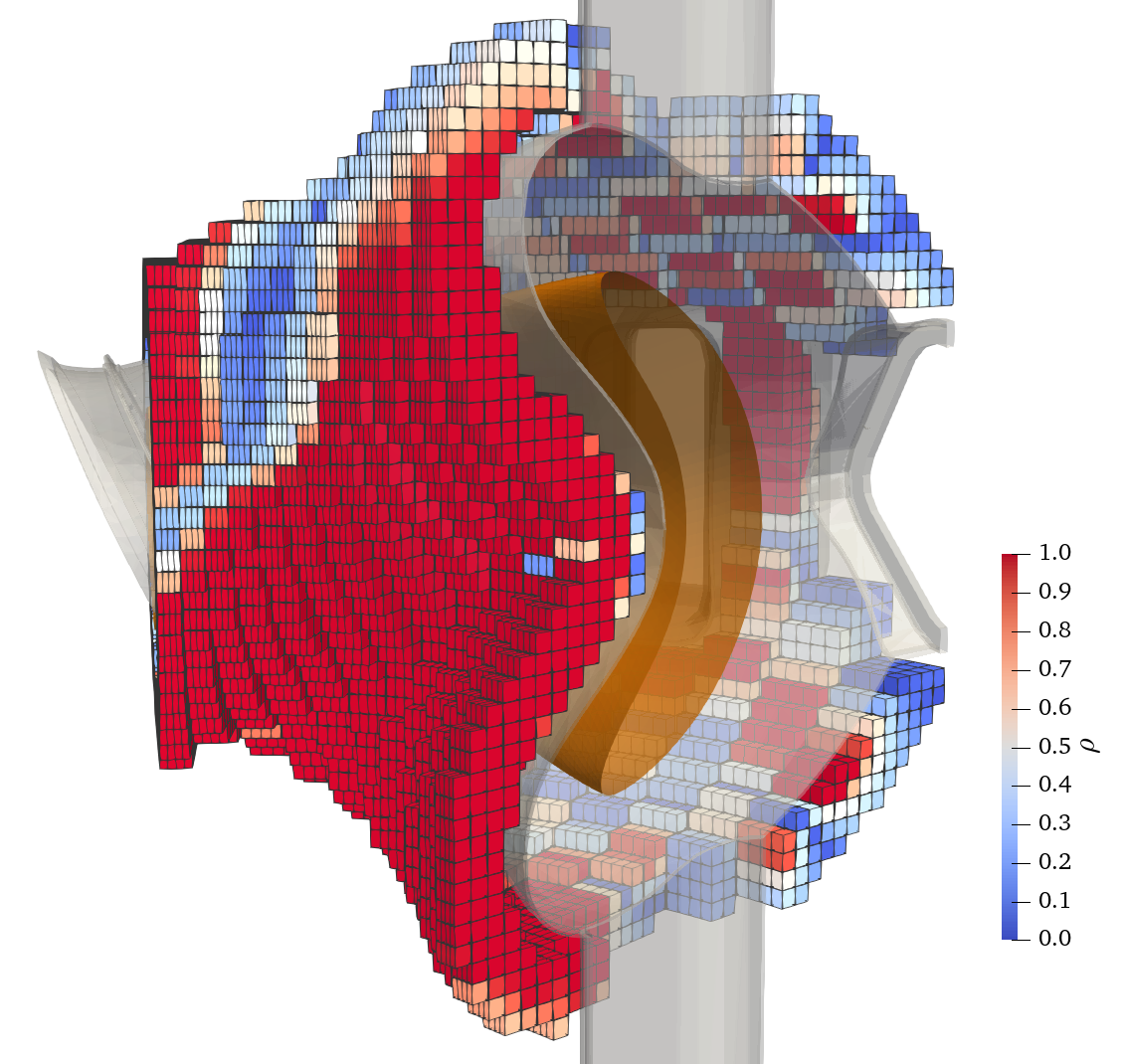}
    \caption{The distribution of the normalized density for the curved bricks case. The magnets are all cuboidal and will be mounted on independent toroidal ribs.}
    \label{fig:trial_24}
\end{figure}

In the above cases, we used structured grids that are discretized along the 
curvilinear coordinates to fit the vacuum vessel.
One can also provide different meshes to meet specific requirements for the 
magnets.
For example, magnet designs with simplified geometry have been 
studied \cite{HammondMagPie} with a view toward reducing the complexity of 
construction.
One of these implementations, which shall be called ``curved bricks'', is 
shown in \Fig{trial_24}.
The magnets are cut along the grids of cylindrical coordinates $(R, \phi, Z)$. 
Each brick has dimensions of
$4.9 \ \mathrm{cm} \times 0.041 \ \mathrm{rad} \times 4.9\ \mathrm{cm}$;
hence, all magnets at a given major radius have the same shape.
The toroidally curved magnets can be effectively approximated by cuboidal magnets.
In \Fig{trial_24}, each magnet is replaced by four cuboidal magnets.
As in previous cases, magnets that collide with selected ports are excluded.
Gaps of 
$0.1 \ \mathrm{cm} \times 0.0025 \ \mathrm{rad} \times 0.1\  \mathrm{cm}$ 
are reserved between each magnet for mounting infrastructure.
The total thickness of the magnet layer is about 20 cm (not uniform).
Details of the design can be found in Ref. \cite{HammondMagPie}.
Now, the magnet information is imported to FAMUS and each magnet is 
approximated by one magnetic dipole located at the center of mass.
The orientations and magnitudes of the dipoles are optimized by FAMUS to 
reduce the normal field error and the number of used magnetic moments. 
The final distribution of the normalized density for each dipole is 
illustrated in \Fig{trial_24}.

The realized normal field error for the radial bricks is 
$1.15 \tento{-6} \ \mathrm{T}^2 \mathrm{m}^2$ and the equivalent volume for 
magnets is $0.64 \  \mathrm{m}^3$.
The normal field error is lower than other cases, which is not surprising 
as the curved-brick case allows the dipole orientation to be varied and 
uses more magnetic moment than the previous orientation-optimized case.
Like the other two solutions, most of the non-zero magnets in the curved 
bricks solution are located at the inboard side; therefore, it may be possible
to refine the configuration by removing many of the outboard magnets altogether.

\subsection{Free-boundary equilibria using permanent magnets}
In the above, we have demonstrated three different designs using permanent magnets.
They are designed following different goals and have various performance in $F_B$ and $F_M$.
In \Tab{comparison}, we list the normal field error evaluated in the form of the mean value of $\abs{\vect{B} \cdot \vect{n}} / \abs{\vect{B}}$ on the plasma boundary and the equivalent volume of NdFeB magnets per half period (calculated by $\sum_{i=1,D}  \abs{\vect{m}}_i / M_0$, where $M_0 = B_r/\mu_0$ and $B_r$ is 1.4 T).
The linear method solution is also included as a reference.
The curved bricks case has the lowest field error while the orientation-optimized case uses the minimum amount of magnets.
Orientation in both cases are allowed to be varied which demonstrates that relaxing the orientation can improve the magnetic efficiency.
The linear method solution has the worst performance as it was not fully optimized and only perpendicular magnets were used.

\begin{table}[]
\centering
\begin{tabular}{lcccc}
%{|l|c|c|c|c|}
%\hline
\toprule
Solutions              & Linear method & Perpendicular-only & Orientation-optimized & Curved bricks \\
\midrule
Ave. $\abs{\vect{B} \cdot \vect{n}} / \abs{\vect{B}}$ & $1.10\tento{-2}$       & $2.46\tento{-3}$            & $1.62\tento{-3}$               & $7.26\tento{-4}$           \\
Magnets vol. ($\mathrm{m}^3$) & 0.68          & 0.50               & 0.43                  & 0.64    \\ 
\bottomrule
\end{tabular}
\caption{Comparisons of the normal field error and the magnet volume for different NCSX permanent magnet designs.}
\label{tab:comparison}
\end{table}

\begin{figure}[hbt!]
\centering
\begin{minipage}{.48\textwidth}
  \centering
  \includegraphics[width=\linewidth]{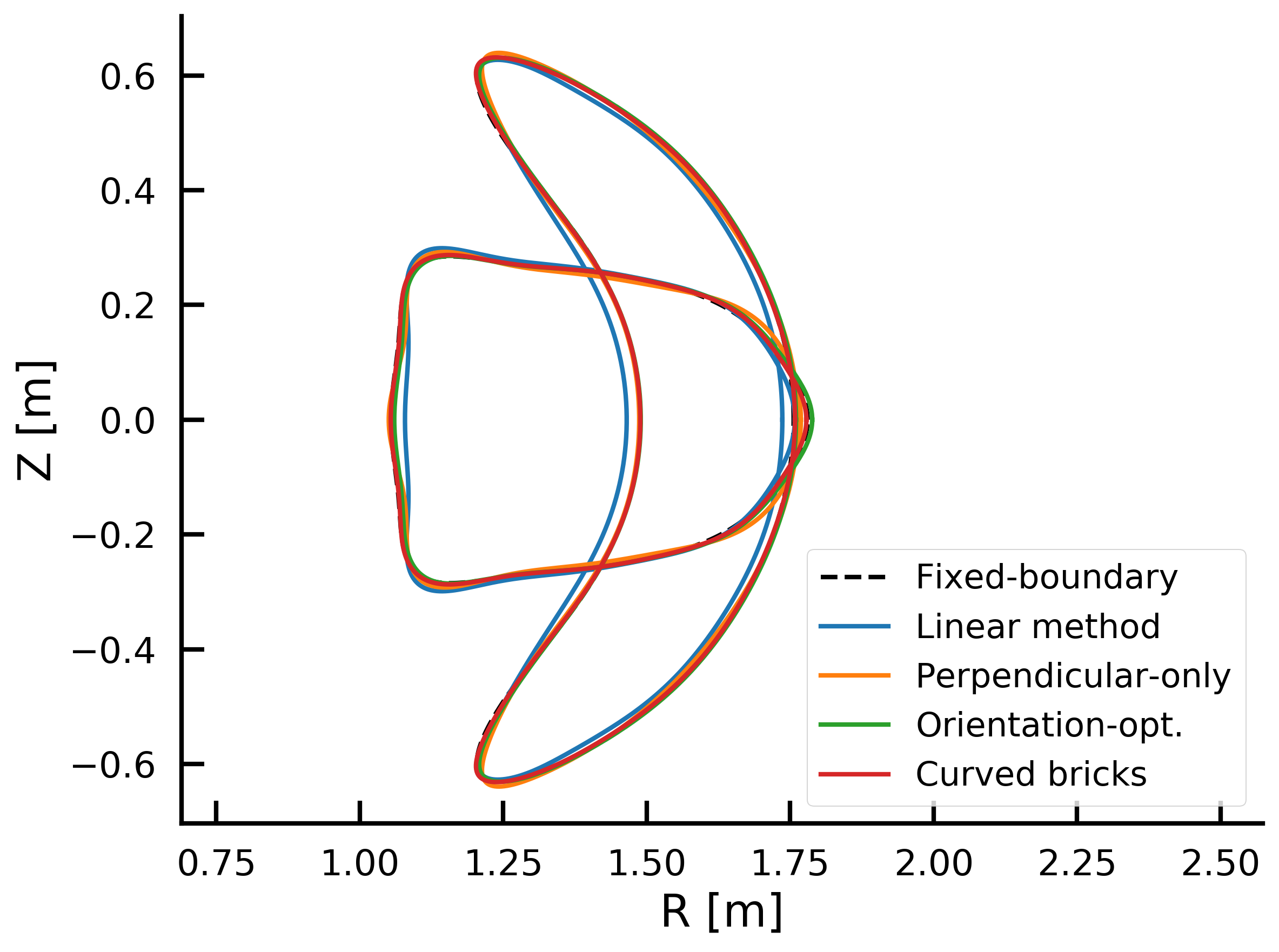}
  \captionof{figure}{LCFS of the free-boundary equilibria from different permanent magnet designs and the fixed-boundary equilibrium. The bean-shaped surfaces locate at $\zeta=0$ and bullet-shaped surfaces at $\zeta=\pi/3$. }
  \label{fig:VMEC_fb}
\end{minipage} \hfill
\begin{minipage}{.48\textwidth}
  \centering
  \includegraphics[width=\linewidth]{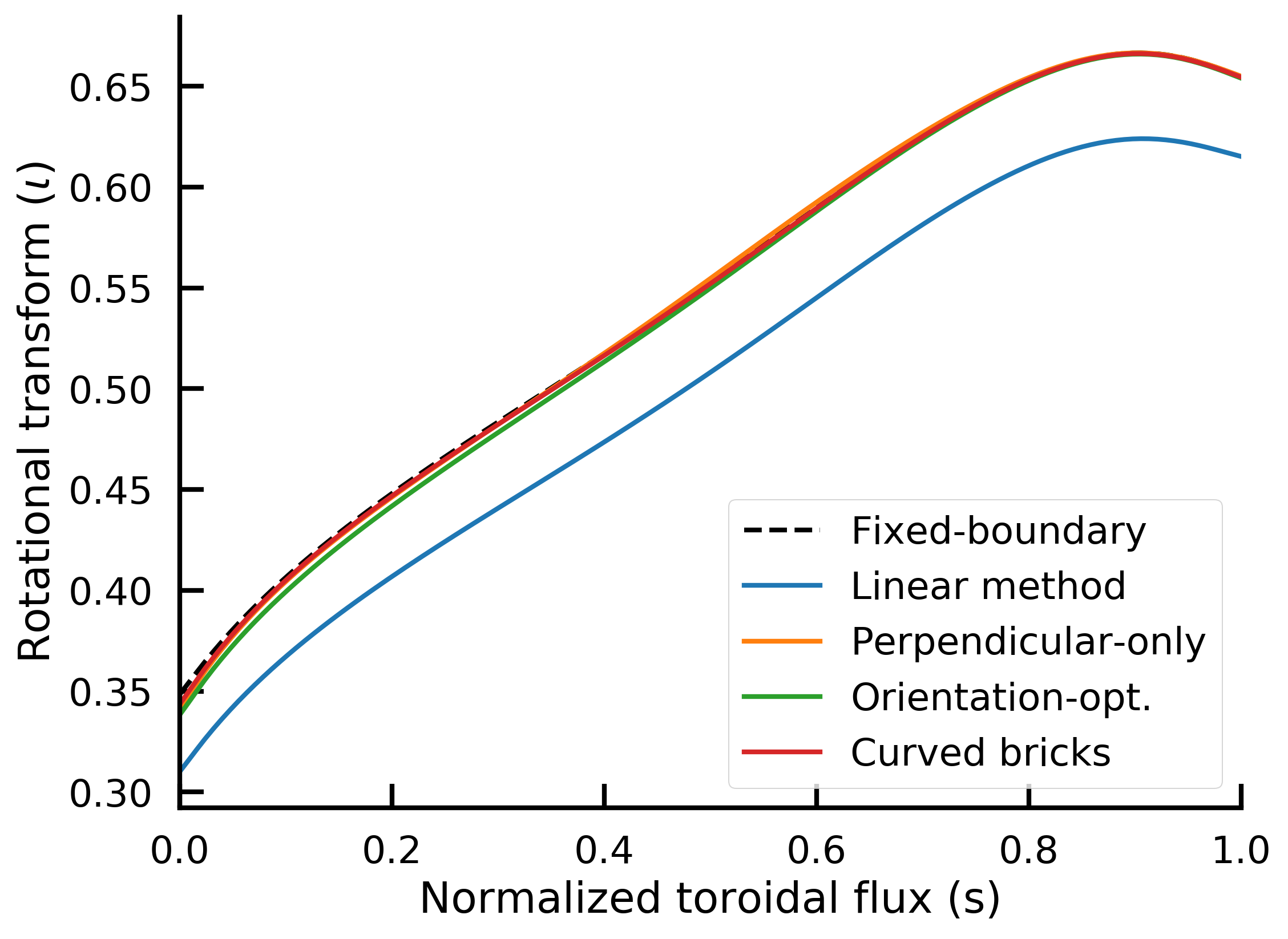}
  \captionof{figure}{Rotational transform profiles of the free-boundary equilibria from different permanent magnet designs and the fixed-boundary equilibrium.}
  \label{fig:iota}
\end{minipage}
\end{figure}

The number of normal field error is one criteria to assess the accuracy of produced magnetic field.
We have also reconstructed the free-boundary equilibria using the VMEC code \cite{VMEC_fb}.
MGRID files are generated (with the same resolution) from the above permanent magnets together with the TF coils.
As shown in \Fig{VMEC_fb}, all the designs have considerably high accuracy in reconstructing the MHD equilibrium compared to the fixed-boundary calculation.
This is also revealed in the comparison of the rotational transform ($\iotabar$) profiles in \Fig{iota}.
The linear method solution has the largest deviations as a reflection of the worst normal field error.
Other solutions have considerably high accuracy.
There are no distinct 
Compared to the fixed-boundary calculations, the discrepancy in the boundary shapes and $\iotabar$ profiles is almost indistinguishable.

\begin{figure}
    \centering
    \includegraphics[width=0.6\linewidth]{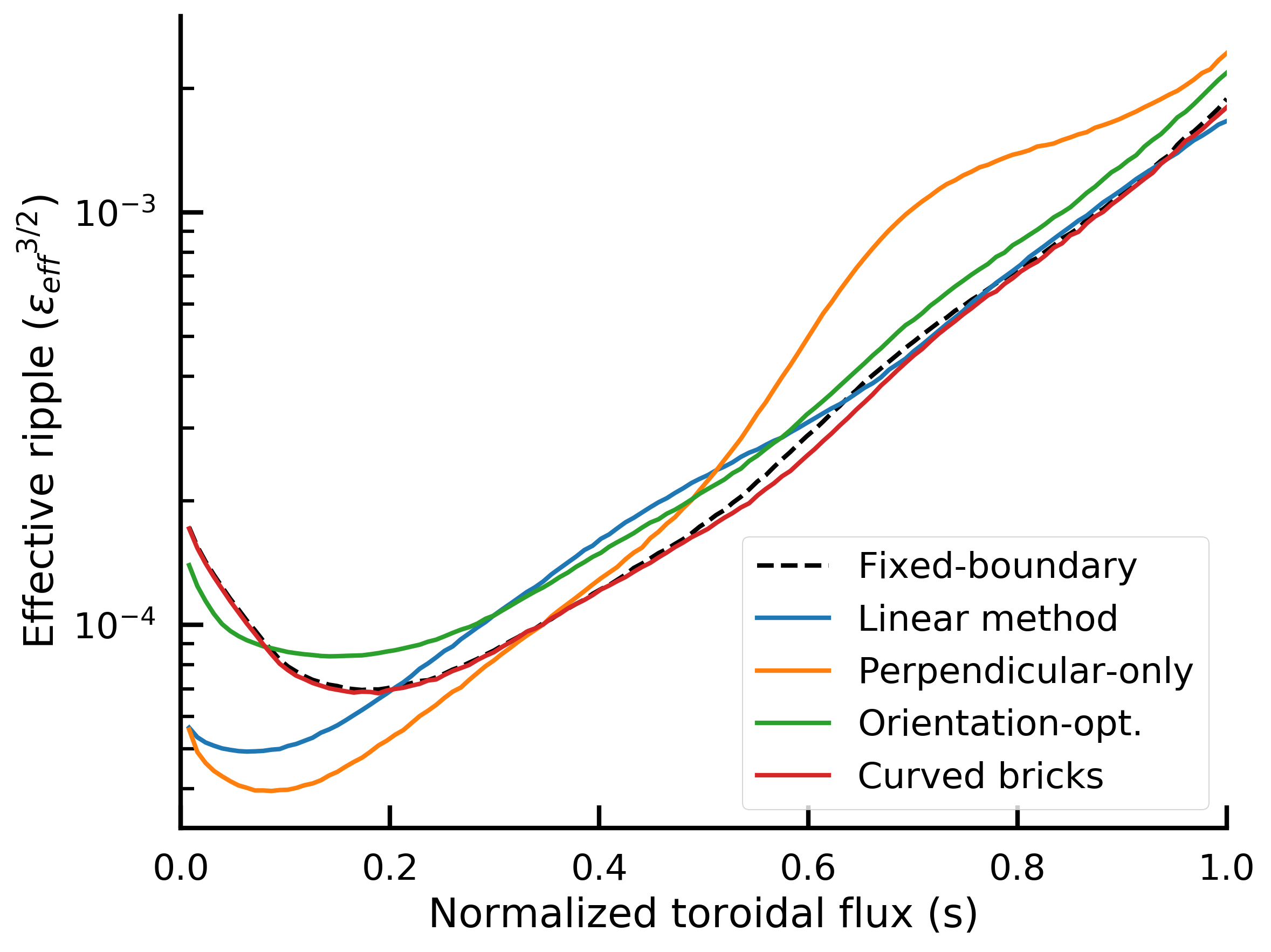}
    \caption{Profiles of ${\epsilon_{eff}}^{3/2}$ for the free-boundary equilibria from different permanent magnet designs and the fixed-boundary equilibrium. All the equilibria are calculated by VMEC with the same resolution.}
    \label{fig:ripple}
\end{figure}

When evaluating the configuration performance, what really matters are the plasma properties.
For example, NCSX was optimized to be quasi-axisymmetric to reduce the neoclassical transport.
Other than comparing the LCFS and $\iotabar$ profiles, we can also check the neoclassical transport of each equilibrium.
Here, we use the NEO code \cite{NEO} to calculate the effective ripple, ${\epsilon_{eff}}$.
The value of ${\epsilon_{eff}}^{3/2}$ of each equilibrium is shown in \Fig{ripple}.
They all have equivalent levels of ${\epsilon_{eff}}^{3/2}$.
The curved bricks case, which has the lowest normal field error, achieves the closest performance to the fixed-boundary equilibrium.
But the linear method case, which has the highest normal field error, is not the most deviated one.
The optimized perpendicular case attains the lowest  ${\epsilon_{eff}}^{3/2}$ at $s=0 \sim 0.35$ and the highest at $s=0.5 \sim 0.9$  ($s$ denotes the normalized toroidal flux).
Good accuracy in the normal field error usually leads to acceptable fidelity in generating the target magnetic field, but it is not completely correlated to physics properties, like the effective ripple shown here.
%negligible differences in normal field errors might cause sizable discrepancy in physics performances.   

%One should note that the hexahedral solution is just one of the possible designs.
%Actually, the normal field error of the hexahedral design can be reduced to the order of $10^{-7}$ when allowing the orientation to be varied.
%It is expected that a better balance between physics accuracy and engineering simplicity can be achieved if more designs are explored.
%More importantly, the target equilibrium of a permanent magnet stellarator, if there is one, needs further optimization to be more suitable for permanent magnets, while in this case we just use an existing equilibrium for demonstrations.
%However, the effective ripple of the hexahedral solution is below 1\%, the level of W7-X \cite{SpongConfigurations}.

\section{Summary} \label{summary}
In this paper we have introduced a topology optimization method to design permanent magnets for stellarators.
The allowable design space for magnets is divided into numerous elements and we use magnetic dipoles to approximate the magnets in each element.
Density method with a penalization technique is used to determine if the magnet should be present in each element.
The FAMUS code is then developed and it can design engineering-feasible permanent magnets for general stellarators satisfying the constraints of the maximum material magnetization and explicit forbidden regions.
We have successfully validated FAMUS with the previously proposed linear method.
Three different permanent magnet designs for a half-Tesla quasi-axisymmetric configuration have been demonstrated.
FAMUS optimized solutions are found to have the following advantages:
\begin{itemize}
    \item[1.] For the perpendicular-only case, the FAMUS solution has lower normal field error and uses less magnets than the linear method.
    \item[2.] FAMUS can optimize the magnitude and/or orientation of all the magnets. By relaxing the orientation of permanent magnets, the so-called ``Halbach'' arrays are formed and the magnetic efficiency is enhanced.
    \item[3.] FAMUS can effectively determine the magnet layout and reserve space for ports. The solutions tend to have completely open outboard access.
\end{itemize}
Free-boundary reconstructions using permanent magnets and planar TF coils are carried out.
All the three designs are proven to have good accuracy in supporting the target equilibrium.

The three permanent magnet designs for NCSX show attractively large access on the outboard side.
This is extremely favorable for future fusion reactors which requires easy access for remote maintenance.
Possible reasons for the outboard clearance are as follows.
The outboard side usually has lower field strength and tends to require less magnets.
Particularly, NCSX has a bean-shaped cross-section.
The concave regions, which requires extra shaping magnetic field, are on the inboard side.
In addition, the penalization on the total amount of magnetic moment tends to remove the magnets on the outboard side as they are further from the plasma and have larger magnetic moment (because the element size is larger).
Although we have only shown the results of a quasi-axisymmetric stellarator, the new tool we have implemented can be applied to any configurations.

In this paper, we are concentrating on the introduction of the new topology optimization method for design permanent magnets.
For future work, we can use the FAMUS code to investigate the difference in the permanent magnet layouts for various stellarators and to study the availability of permanent magnets in reactor-size experiments.
We can also adopt more advanced optimization algorithms, like the stochastic gradient descent \cite{SGD}, to better handle the local minimum.
More engineering constraints, like the calculation of magnetic forces and the assembly tolerance, could also be implemented in the future.
FAMUS requires non-trivial efforts on generating meshes, which is particularly challenging when dealing with the complicated geometries of stellarators.
Powerful meshing tools can be used in the future to help obtain better solutions.

%We can also do discrete optimization for the orientation.
%In reality, a simple and efficient design should use a combination of limited orientations.

% dependence on mesh
% local minimum
% discrete optimization

% 3. Potential future work
% dimensionality reuction, surrogate model, stochastic gradient, machine learning 

\section*{Acknowledgments}
% CZ gratefully appreciates fruitful discussions with xxx.
This work was supported by the U.S. Department of Energy under Contract No. DE-AC02-09CH11466 through the Princeton Plasma Physics Laboratory.

\begin{appendices}
\numberwithin{equation}{section}

\section{Evidence of local minima} \label{local_minimum}
\begin{figure}
    \centering
    \includegraphics[width=0.8\linewidth]{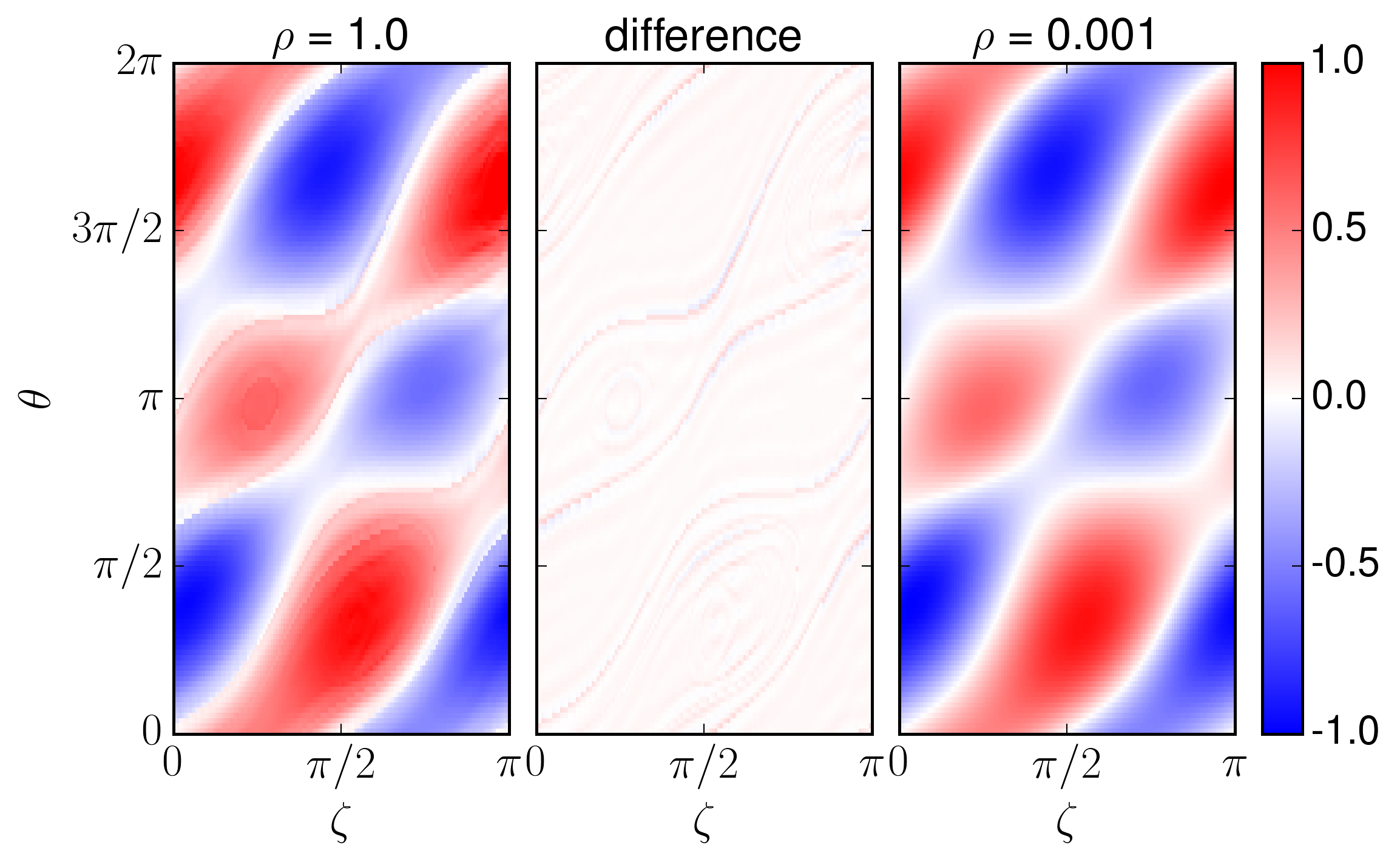}
    \caption{Distributions of two FAMUS runs for the rotating ellipse case initialized with different densities ($\rho=1.0$ and $\rho=0.001$ ). The $\rho=0.001$ case (right) is the one shown in \Fig{surf_mag_comp}.}
    \label{fig:local_min}
\end{figure}

For the benchmark case, we initialized the FAMUS run with a uniform normalized density, $\rho = 0.001$.
However, if the initial guess was $\rho = 1.0$, we would get a different solution, as shown in \Fig{local_min}.
The normal field error is $F_B = 3.82 \tento{-11}\ \mathrm{T}^2 \mathrm{m}^2$ after 500 iterations.
Although the normal field error is considerably low, the distributions of the normalized density for the two solutions are distinct.
The differences in the two solutions initialized from different guesses indicate that they are trapped in different local minima.
This is not surprising because there are 16384 variables in the optimization problem and we are not using any global optimization algorithms.

\begin{figure}
    \centering
    \includegraphics[width=0.8\linewidth]{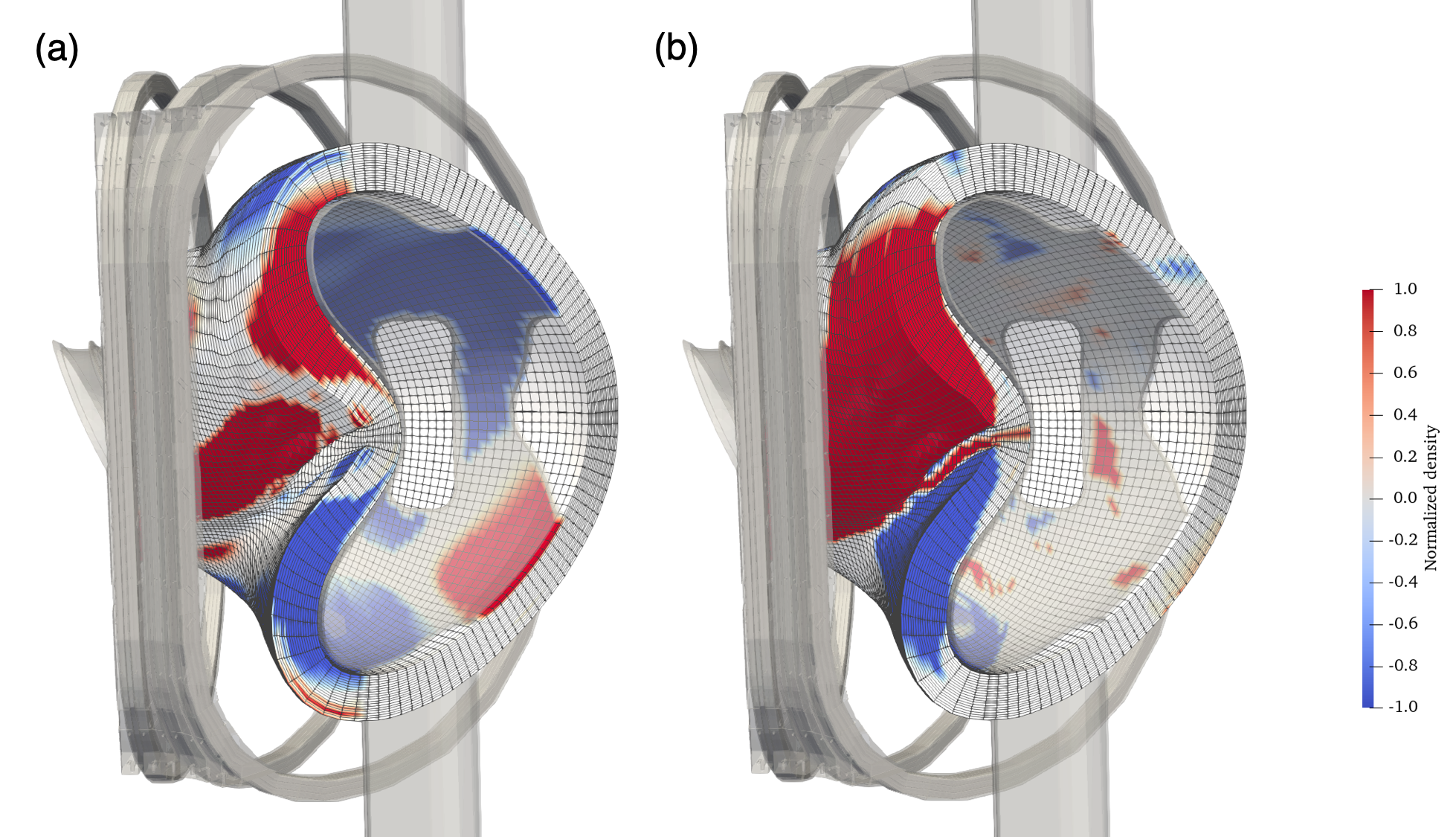}
    \caption{The distributions of the normalized density for two perpendicular-only permanent magnet designs for the half-Tesla NCSX. The two solutions were optimized by FAMUS with the same settings, while the left one (a) was initialized with an uniform guess of $\rho=0.001$ and the right one (b) was $\rho=1.0$. }
    \label{fig:ncsx_local_min}
\end{figure}

The clear discrepancy in FAMUS solutions from different initial guesses can also be observed in \Fig{ncsx_local_min}.
Two different perpendicular-only solutions, as introduced in section \ref{perp_only}, have been obtained by using FAMUS with the same settings.
The only difference is that one was initialized with $\rho=0.001$ and the other one with $\rho=1.0$.

\section{Least-squares minimization to calculate the magnetic moment} \label{least_square}
\Eqn{dipole} implies that the magnetic field is linear to magnetic moment, if we fix the position of each dipole.
We want to minimize the normal field error on the plasma boundary, $F_B$, which is quadratic of $\vect{B}_{M}$.
It can be converted to a least-squares minimization problem and solved linearly.
The derivative of $F_B$ with respect to an arbitrary parameter, for instance $m_k^i$ ($i$-th component of the magnetic moment of the $k$-th dipole), is
\begin{align}
    \pdv{F_B}{m_k^i} & = \iint_S 2 \left ( \vect{B}_{M} \cdot \vect{n} - B_n^{tgt} \right ) \left (\pdv{\vect{B}_{M}}{m_k^i} \cdot \vect{n} \right) \dd{a} \ , \nonumber \\ 
    & = \iint_S 2 \left [  \frac{\mu_0}{4\pi} \sum_{j=1}^{D} \left (\frac{3 \vect{m}_j \cdot \vect{r}_j}{|\vect{r}_j|^5} \vect{r}_j - \frac{1}{|\vect{r}_j|^3} \vect{m}_j \right) \cdot \vect{n} - B_n^{tgt} \right ] 
    \left [ \frac{\mu_0}{4\pi} \left (\frac{3 \vect{r}_k \cdot \vect{n}}{|\vect{r}_k|^5} r_k^i - \frac{1}{|\vect{r}_k|^3} n^i \right) \right ] \dd{a} \ , \nonumber  \\
    %& = \iint_S 2 \left [  \frac{\mu_0}{4\pi} \sum_{j=1}^{D} \left (\frac{3 \vect{r}_j \cdot \vect{n} }{|\vect{r}_j|^5} \vect{r}_j - \frac{1}{|\vect{r}_j|^3} \vect{n} \right) \cdot \vect{m}_j - B_n^{tgt} \right ] 
    %\left [ \frac{\mu_0}{4\pi} \left (\frac{3 \vect{r}_k \cdot \vect{n}}{|\vect{r}_k|^5} r_k^i - \frac{1}{|\vect{r}_k|^3} n^i \right) \right ] \dd{a} \ , \nonumber \\   
    %& = \frac{\mu_0}{2\pi} \Delta \t \Delta \z \sum_{i_{\t}} \sum_{i_{\z}} \left [  \frac{\mu_0}{4\pi} \sum_{j=1}^{D} \left (\frac{3 \vect{r}_j \cdot \vect{n} }{|\vect{r}_j|^5} \vect{r}_j - \frac{1}{|\vect{r}_j|^3} \vect{n} \right) \cdot \vect{m}_j - B_n^{tgt} \right ] 
    %\left (\frac{3 \vect{r}_k \cdot \vect{N}}{|\vect{r}_k|^5} r_k^i - \frac{1}{|\vect{r}_k|^3} N^i \right)  \ , \nonumber \\ 
    & = 2 \Delta \t \Delta \z \sum_{i_{\t}} \sum_{i_{\z}} \left [  \frac{\mu_0}{4\pi} \frac{1}{N} \sum_{j=1}^{D} \left (\frac{3 \vect{r}_j \cdot \vect{N} }{|\vect{r}_j|^5} \vect{r}_j - \frac{1}{|\vect{r}_j|^3} \vect{N} \right) \cdot \vect{m}_j - B_n^{tgt} \right ] 
    \left [ \frac{\mu_0}{4\pi} \left (\frac{3 \vect{r}_k \cdot \vect{N}}{|\vect{r}_k|^5} \vect{r}_k - \frac{1}{|\vect{r}_k|^3} \vect{N}  \right) \cdot \vect{e_i} \right ] \ , \nonumber \\
    & = 2 \Delta \t \Delta \z \sum_{i_{\t}} \sum_{i_{\z}} \left (  \frac{1}{N} \sum_{j=1}^{D} \vect{g}_j \cdot \vect{m}_j - B_n^{tgt} \right ) 
    \left ( \vect{g}_k  \cdot \vect{e_i} \right ) \ ,   
\end{align}
where $\vect{g} (\t, \z)$ is the ``inductance'' matrix as calculated by \Eqn{inductance}.
%where $ \ds \vect{g}_k (\t, \z) = \frac{\mu_0}{4\pi} \left (\frac{3 \vect{r}_k \cdot \vect{N}}{|\vect{r}_k|^5} \vect{r}_k - \frac{1}{|\vect{r}_k|^3} \vect{N}  \right)$ is similar to the inductance matrix in Eq. (A.10) of REGCOIL.
%(m_p^x r_p^x + m_p^y r_p^y + m_p^z r_p^z)  (m_p^x n^x + m_p^y n^y + m_p^z n^z) 
To minimize $F_B$, we can solve $\partial F_B / \partial m_k^i = 0 $, and we obtain a linear system,
\begin{equation}
    \sum_{j} \sum_{l} A^{i,l}_{j,k} m_j^l = b_{k}^i \ ,
\end{equation}
where the matrix element is 
\begin{equation} \label{eq:A_B}
    A^{i,l}_{j,k} = \Delta \t \Delta \z \sum_{i_{\t}} \sum_{i_{\z}} \left (  \frac{1}{N} {g}^l_j {g}^i_k \right ) \ ,
\end{equation}
and the right-hand side 
\begin{equation} \label{eq:b_B}
    b_{k}^i = \Delta \t \Delta \z \sum_{i_{\t}} \sum_{i_{\z}} \left ( B_n^{tgt} {g}^i_k \right ) \ .
\end{equation}

We can also add a regularization term, $F_M$, which is the overall magnetic moment (equivalent to the total amount of PM material).
The combined objective function, $F $, is now expressed as \Eqn{target}.
Likewise, if we want to minimize $F_M$ only, we have
\begin{equation}  \label{eq:Ab_M}
    \pdv{F_M}{m_k^i} = 2 m_k^i = 0 \ .
\end{equation}

After packing all the parameters (magnetic moments) into one column vector with $3D$ entries, $\vect{x} = [m_1^x, m_1^y, m_1^z, m_2^x, \cdots , m_D^z]^T$, we can solve the linear equations,
\begin{equation}
    \vect{A} \vect{x}  = \vect{b} \ ,
\end{equation}
to find the minimum of the combined objective function, $\partial F / \partial \vect{x} = 0$.
The size of the matrix $\vect{A}$ is $3D \times 3D$, while $\vect{b}$ is a $3D \times 1$ column vector.
The components of $\vect{A}$ and $\vect{b}$ are computed using \Eqn{A_B}, \Eqn{b_B} and \Eqn{Ab_M}.

If we separate the orientation and magnitude of each dipole, the magnetic moment can be represented as $\vect{m}_j = I_j \hat{\vect{m}}_j$.
Similarly, we can write down the linear equations for solving the magnitude,
\begin{equation}
    \sum_{j} A_{j,k} I_j = b_{k} \ ,
\end{equation}
where 
\begin{equation}
    A_{j,k} = \Delta \t \Delta \z \sum_{i_{\t}} \sum_{i_{\z}} \left (  \frac{1}{N} h_j h_k \right ) \ ,
\end{equation}
\begin{equation}
    b_{k} = \Delta \t \Delta \z \sum_{i_{\t}} \sum_{i_{\z}} \left ( B_n^{tgt} h_k \right ) \ .
\end{equation}
Right now, the ``inductance'' equation is
\begin{equation}
    h_k(\t,\z) = \frac{\mu_0}{4\pi} \left (\frac{3 \vect{r}_k \cdot \vect{N}}{|\vect{r}_k|^5} \vect{r}_k \cdot \hat{\vect{m}}_k - \frac{1}{|\vect{r}_k|^3} \vect{N} \cdot \hat{\vect{m}}_k \right)
\end{equation}

Both the two least-squares minimization methods cannot explicitly address the constraint of the maximum allowable magnetization.
When the number of dipoles is large, inverting a large matrix using the normal QR decomposition or singular value decomposition might be considerably slow.

% \section{Non-linear optimization}
% \begin{equation}
%     \vect{m} = \rho {M_0} V \hat{\vect{e}}_m
% \end{equation}

% \begin{equation}
%     \hat{\vect{e}}_m = \{ \sin{\t} \cos{\phi}, \sin{\t} \sin{\phi},  \cos{\t}\}
% \end{equation}

% \begin{equation}
%     \rho = p^q, p \in [-1,1]
% \end{equation}

% \begin{equation}
%   \ds \min_{p} V_{eff} = \sum_i^{N} \rho 
% \end{equation}

% \begin{equation}
%   \ds \mathcal{F}[p, \t, \phi] = \oint_S \half (\vect{B} \cdot \vect{n})^2 \dd{S} + \lambda \sum_i^{N} \rho_i
% \end{equation}

% $\curl{\vect{M}}=0 \; \text{in} \; \Omega$
% \begin{equation}
%     \vect{M} =
%     \begin{cases}
%     -\grad{\Phi} \quad \text{, on}\; \partial \Omega_{inner} \\
%     \quad 0  \quad \quad \text{, on}\; \partial \Omega_{outer} 
%     \end{cases}
% \end{equation}

\end{appendices}

\section*{References}
\bibliographystyle{unsrt}
\bibliography{topo_opt}

\end{document}

%% file: packages_commands.tex
 \usepackage{color}
 \usepackage[utf8]{inputenc}
 \usepackage{bm}
 \usepackage{graphicx, subcaption}
 \usepackage{latexsym}
 \usepackage{pifont}
 \usepackage{amssymb}
 \usepackage{hyperref}
 \usepackage{titlesec}
 \expandafter\let\csname equation*\endcsname\relax
 \expandafter\let\csname endequation*\endcsname\relax
 \usepackage{amsmath}
 \usepackage{physics}
 \usepackage{listings}
 \usepackage{verbatim}
 \usepackage{geometry}
 \usepackage{booktabs}
 \usepackage{enumitem}
 \usepackage[normalem]{ulem} % for sout;
 \usepackage{breqn}
 \usepackage[toc,page]{appendix}
 
 \definecolor{red}{rgb}{1.0,0.0,0.0}
 \definecolor{gre}{rgb}{0.0,1.0,0.0}
 \definecolor{blu}{rgb}{0.0,0.0,1.0}

 \definecolor{ora}{rgb}{1.0,0.5,0.0}
 \definecolor{gra}{rgb}{0.0,0.5,1.0}

 \newcommand{\ba}{\begin{abstract}}
 \newcommand{\ea}{\end{abstract}}
 \newcommand{\be}{\begin{equation}}
 \newcommand{\ee}{\end{equation}}

 \newcommand{\rom}[1]{\uppercase\expandafter{\romannumeral #1 \relax}}
 
 \newcommand{\iotabar}{\mbox{$\,\iota\!\!$-}}
 \newcommand{\Fig}[1]{Fig.\ref{fig:#1}}
 \newcommand{\Tab}[1]{Table.(\ref{tab:#1})}
 \newcommand{\Eqn}[1]{Eq.(\ref{eq:#1})}
 
 \renewcommand{\t}{\theta}
 \newcommand{\z}{\zeta} 
 \newcommand{\ds}{\displaystyle}
 
 \newcommand{\vect}[1]{\mathbf{#1}}

 \newcommand{\tento}[1]{\times 10^{#1}}